\documentclass[twocolumn]{aastex7}
\usepackage{mathtools}

\begin{document}

%\title{Complex Interactions of Thermal Evolution, Mass Loss, and Hydrogen Dissolution in Sub-Neptune Magma Oceans}

\title{Hydrogen and Helium Dissolution, Outgassing, and Loss in Evolving Sub-Neptune Magma Oceans: \\ Examining Demographic Features and Radius Evolution}

\author[0000-0003-3980-7808]{Yao Tang}
\affiliation{Department of Astronomy and Astrophysics, University of California, Santa Cruz \\
1156 High Street, Santa Cruz, CA 95064, USA}
\email{yaotang@ucsc.edu}

\author[0000-0002-9843-4354]{Jonathan J. Fortney}
\affiliation{Department of Astronomy and Astrophysics, University of California, Santa Cruz \\
1156 High Street, Santa Cruz, CA 95064, USA}
\email{jfortney@ucsc.edu}

\author[0000-0003-2915-5025]{Laura K. Schaefer}
\affiliation{Department of Earth and Planetary Sciences, Stanford University, Stanford, CA 94305, USA}
\email{lkschaef@stanford.edu}

\author[0000-0002-8518-9601]{Peter Gao}
\affiliation{Earth and Planets Laboratory, Carnegie Institution for Science, Washington DC, USA}
\email{pgao@carnegiescience.edu}

\begin{abstract}
Sub-Neptunes' molten interiors are expected to accommodate large quantities of volatiles, potentially altering their radius evolution. Previous studies have examined this effect in isolation with simplified evolution modeling, often assuming idealized interior and atmospheric conditions. To address this limitation, we introduce SEAMIST, a unified evolution model for sub-Neptunes and super-Earths that self-consistently combines, for the first time, interior structure, cooling, rock/iron solidification, boil-off, photoevaporation, H/He dissolution, and atmospheric composition. SEAMIST considers both a partially soluble case, in which hydrogen partitions into the magma ocean with a fraction set by mantle-envelope boundary conditions, and a fully miscible case, in which hydrogen may fully dissolve into the magma ocean at high temperatures. We identify a novel catastrophic boil-off mechanism, triggered by a positive feedback between hydrogen outgassing and mass loss that can operate billions of years after disk dispersal. In partially soluble models, the impact of H/He dissolution on radius evolution is modest. This contrasts with previous expectations, as we find that increasing hydrogen abundance from outgassing enhances mass-loss efficiency, counterbalancing volatile replenishment from the rock/iron interior. Fully miscible hydrogen, in contrast, enhances envelope survival around low-mass stars. Overall, at intermediate to low masses, mass-radius curves from partially soluble models match observed distributions. The fully miscible case predicts a radius peak and excess planets around low stellar masses that appear inconsistent with current observations, although it reproduces the observed radius ``cliff" near 4$R_\oplus$ at higher masses. Our results suggest that high metallicity may explain the cliff, although alternatives cannot be ruled out.
\end{abstract}

\keywords{Planet interior --- Planet atmosphere}

\section{Introduction} \label{sec:intro}
A remarkable finding of the Kepler mission is the bimodal radius distribution \citep{Fulton17} of close-in, small exoplanets, the most common planets discovered in the galaxy so far \citep{Borucki11,Batalha12}. These planets are separated into two distinct populations by a radius gap. Planets with larger radii ($\sim$2-4 $R_\oplus$) are inferred to possess H/He envelopes that produce low bulk densities \citep{Rogers10,Lopez14} and are referred to as sub-Neptunes, whereas the higher densities of smaller planets ($\sim$1-1.5 $R_\oplus$) are consistent with rocky, Earth-like compositions \citep{Zeng19} and are commonly referred to as super-Earths.

One widely discussed hypothesis proposes \citep{Lopez12,Owen13,Jin14,Ginzburg18,Gupta20,Affolter23} that some sub-Neptunes retain volatile envelopes accreted from the protoplanetary nebula against atmospheric escape, whereas others undergo complete loss of their initial envelope mass, forming super-Earths. This evolutionary divergence is primarily regulated by planetary mass and incident bolometric flux \citep{Lopez13,Owen17,Ginzburg18,Tang24}: lower-mass planets at higher irradiation are more prone to envelope loss because of their shallower gravitational potential wells and the greater energy available to drive atmospheric winds. 

Two escape mechanisms have been widely discussed over the past decade to explain the radius gap: photoevaporation and core-powered mass loss \citep{Ginzburg16}. Both mechanisms successfully reproduce the observed planet occurrence as a function of radius. However, significant uncertainties remain. These models rely on substantially different initial conditions and core-mass distributions \citep{Owen17,Ginzburg18,Rogers21}, and they are tuned to reproduce the observed radius distribution in the mass space, struggling to reproduce observations in other parameter spaces, such as orbital period and stellar mass \citep{Dattilo23}. This degeneracy highlights the need for a high-dimensional assessment of the relevant physical processes within a unified framework.

More recently, a critical reassessment \citep{Tang24} challenged the physical viability of core-powered mass loss, arguing that core luminosity plays only a limited role in setting the mass-loss rate for bolometric-energy-driven escape. That work instead favors a rapid escape mechanism known as ``boil-off'' \citep{OwenWu16}, which operates under rapid disk-dispersal conditions rather than Gyr timescales found in \citet{Ginzburg18}. Boil-off can efficiently strip the entire envelope of low-mass, highly irradiated planets and is expected to dominate the cumulative mass loss over photoevaporation for H/He-dominated sub-Neptunes \citep{Tang25b}, often referred to as gas dwarfs.

Previous sub-Neptune evolution models commonly coupled photoevaporation with the thermal contraction of the gaseous envelope \citep{Lopez12,Owen13,Jin14,Howe15,Chen16,Rogers21}. Thermal contraction is driven by envelope cooling and the associated increase in density. However, in those works, the evolution of the rock/iron core received comparatively little attention. Pioneering work by \citet{Vazan18b} first highlighted the potential importance of core evolution in sub-Neptunes. This was later expanded by the self-consistent mantle-core-envelope evolution model of \citet{Tang25}, which demonstrated that deep, long-lived magma oceans should be ubiquitous in these planets. Under such conditions, volatile dissolution becomes important, as suggested by high-pressure experiments of \citet{Hirschmann12}, modifying the envelope's thermal evolution and potentially hindering the transformation into super-Earths by buffering photoevaporative mass loss \citep{Chachan18}. In addition, magma oceans enable active chemical interactions between H-rich envelopes and oxygen-bearing silicate melts \citep{Kite20}.

However, although pioneering, previous models that considered magma-ocean-envelope interactions did not incorporate a number of vital physical processes. Some neglected atmospheric mass loss altogether \citep{Schlichting22,Rogers25}, others assumed fixed atmospheric compositions and therefore ignored compositional evolution \citep{Chachan18,Kite19,Rogers25}, and most did not account for coupling with the deeper iron core. In contrast, first-principle simulations reveal a substantially higher hydrogen solubility in the liquid iron than magma ocean \citep{Tagawa21,Luo24,Gupta25}. Additionally, recent \textit{ab initio} molecular dynamics simulations suggest the possibility of fully miscible hydrogen in silicate \citep{Stixrude25}, in which case a large fraction, potentially most, of the initial hydrogen inventory may be dissolved into the mantle. These uncertainties directly impact radius evolution. 

The evolution of atmospheric composition arising from the dissolution process is expected to play a critical role in setting planetary radii. As shown by \citet{Tang25}, the radiative atmosphere can account for a substantial fraction of the apparent radius of low-mass sub-Neptunes, making radius evolution highly sensitive to compositional changes. Moreover, the mass loss rate sensitively depends on the lower boundary conditions, and are expected to be tightly coupled to those processes.

Earlier models generally neglected solidification, assuming a fully molten mantle, despite the fact that the melt fraction strongly controls volatile storage, with solid silicates having much lower volatile storage capacities. Furthermore, the early ``boil-off'' phase was largely ignored, even though rapid decompression at the mantle-envelope boundary \citep{Tang24} should strongly favor volatile outgassing during this stage. Moreover, the transition from sub-Neptunes to super-Earths has often been treated as an ambiguous boundary in population and evolutionary modeling. In many studies, planets are considered atmosphereless once the envelope mass fraction falls below a certain threshold \citep[e.g., $10^{-6}$ in ][]{Ginzburg18}, and the subsequent evolution into a super-Earth is not explicitly modeled. However, \citet{Tang25} demonstrated that this critical envelope mass fraction strongly depends on planetary mass and incident flux and showed that this transformation boundary typically occurs at a much higher mass fraction, $\sim10^{-4}$.

Motivated by these limitations, we develop a unified framework that self-consistently couples boil-off, photoevaporation, thermal evolution, rock/iron solidification, volatile dissolution, and atmospheric compositional evolution for the first time, building upon \citet{Tang25}. Compared to the framework of \citet{Tang25}, this framework bridges sub-Neptune and terrestrial planet evolution, seamlessly combining their evolution and enabling future extensions to include secondary atmosphere formation. This framework also incorporates new modules for H/He dissolution, miscibility and outgassing, and extends the numerical scheme to model the resulting compositional evolution of the interior We term this new framework Synthetic Evolution and Atmosphere Model In Sub-Neptunes and Terrestrial planets (SEAMIST). We self-consistently incorporate both partially soluble and fully miscible cases, the latter representing an end-member scenario, into our coupled interior-composition-atmosphere-escape evolution model. In this study, we focus primarily on the effects of H/He dissolution rather than chemical reactions and fractionation of the outflow.

In addition to the radius gap, another intriguing demographic feature is the apparent planetary radius cutoff at $\sim4R_\oplus$ in the sub-Neptune population, known as the ``radius cliff''. The origin of this radius cliff has remained puzzling, as standard atmospheric mass-loss models alone cannot readily reproduce it \citep{Dattilo24}. Volatile dissolution into a molten interior has been proposed as one possible explanation \citep{Kite19}. An alternative hypothesis is that the observed sub-Neptune population reflects high interior metallicities, potentially extending into a water-dominated regime, the so-called ``water worlds'' \citep{Rogers10,Luque22,Aguichine25}. Motivated by these possibilities, we reexamine the origin of the radius cutoff using the SEAMIST framework by systematically exploring the coupled effects of metallicity, volatile dissolution, and atmospheric escape.

The main objectives of this paper are:
\begin{itemize}
\item Introduce the SEAMIST framework, in Section \ref{sec:methods}.
\item Identify the key feedback mechanisms governing the coupled evolution in Section \ref{sec:results}.
\item Investigate the role of H/He solubility, both partially soluble and fully miscible, on planetary radius evolution.
\item Explore the impact of solubility in a high-dimensional parameter space, including metallicity, planetary mass, stellar mass, and orbital separation, and assess its influence on planetary population statistics.
\item Discuss implications for sub-Neptune observation and theory in Section \ref{sec:disc}
\end{itemize}

\section{Model Description} \label{sec:methods}
\subsection{Structure Modeling} \label{subsec:interior}
SEAMIST computes a full structure model at each evolutionary timestep. The structural framework follows \citet{Tang25} and consists of three independently evolving adiabatic interior layers, the iron core, silicate mantle, and convective gaseous envelope, and a radiative atmosphere above the radiative-convective boundary (RCB). The thermal states of these layers are tracked using the core potential temperature $T_c$, the mantle potential temperature $T_m$, and the envelope specific entropy $s$. Between timesteps, the potential temperatures are redefined to the surface pressures of the core and mantle rather than at fixed pressure levels. 

SEAMIST computes the physical structure of each interior layer by solving the hydrostatic equilibrium equations using an iterative relaxation method:
\begin{equation}
\label{masscon}
\frac{dr}{dm} = \frac{1}{4\pi r^2\rho}
\end{equation}

\begin{equation}
\label{hydrostatic}
\frac{dP}{dm} = -\frac{Gm}{4\pi r^4}
\end{equation}
where $r$, $\rho$ and $P$ denote the radius, density, and pressure at a given mass shell, and $m$ is the mass enclosed within that shell.

\subsubsection{Radiative Atmosphere}
During the first few million years, particularly throughout the boil-off phase, the radiative atmosphere is often in a steady-state outflow rather than hydrostatic equilibrium. In such cases, we determine the planetary radius using an isothermal Parker wind model. The later evolution follows a non-isothermal double-stream radiative transfer model \citep{Guillot10}. To ensure a smooth transition between the two regimes, we apply the transition function described in \citet{Tang25b}:
\begin{equation}
\label{activation}
\phi(P,t) = \frac{1}{2} + \frac{1}{2}\tan\left[\frac{\log(v(P,t))-\log(v_{\rm{crit}})}{w_{\rm{log,v}}}\right] \\
\end{equation}
The transition function $\phi(P,t)$ is evaluated at each pressure level $P$ and age $t$. Here, $v_{\rm{crit}}$ is the transition velocity and $w_{\rm{log,v}}$ is the transition width in logarithmic velocity space. All velocities are normalized by the local sound speed. The radius is then computed as:
\begin{equation}
\label{Rtrans}
R(P,t) = \phi R_{s-s}(P,t) + (1-\phi)R_{hydro}(P,t)
\end{equation}
where $R_{s-s}$ is the radius given by the isothermal steady-state solution, and $R_{hydro}$ is the radius from the non-steady-state hydrostatic solution. As demonstrated in \citet{Tang25}, accounting for variable gravity and non-isothermality is essential; neglecting these effects can lead to a significant underestimation of the planetary radius in low-mass planets. We apply Eq. \ref{Rtrans} to evaluate $R(P,t)$ below the photoionization base, or the lower boundary of our hydrodynamic photoevaporation model at 1$\mu$bar. Above this level, the atmospheric structure is determined directly by the photoevaporation model (see Section 2.3).

SEAMIST defines the planetary radius $R_p$ at the optical transit pressure level corresponding to the Kepler bandpass, which is determined using PLATON \citep{Zhang2019}. To facilitate this, we constructed a grid of transit pressures as a function of metallicity, C/O ratio, surface gravity, and incident bolometric flux. For the present study, we adopt a fixed C/O ratio of 0.5.

\subsubsection{Envelope, Mantle and Core}
The thermal and physical structures of the envelope are determined directly from the equation of state (EOS), which is pre-tabulated as a function of $s$ and $P$ for computational efficiency. We model the envelope as an H-He-metal mixture, with metals represented by water. The helium-to-hydrogen mass ratio is fixed to Y/(X+Y)=0.275 across all models. The hydrogen and helium EOSs are taken from \citet{Chabrier19}, and the water EOS follows \citet{Mazevet19}. In sub-Neptunes, the envelope is dominated by molecular hydrogen; however, the adopted EOS self-consistently accounts for hydrogen dissociation and ionization.

The mixture EOS is computed using the additive-volume method. The density $\rho(T,P)$ is then evaluated by:
\begin{equation}\label{addrho}
    \frac{1}{\rho(T,P)} = \frac{X}{\rho_H(T,P)} + \frac{Y}{\rho_{He}(T,P)} + \frac{Z}{\rho_{Met}(T,P)}
\end{equation}
where $\rho(T,P)$, $\rho_{He}(T,P)$ and $\rho_{Met}(T,P)$ are the densities of hydrogen, helium, and metals at the given temperature and pressure. The specific entropy of the mixture is given by:
\begin{equation}\label{addentropy}
    s(T,P) = s_H(T,P) X + s_{He}(T,P) Y + s_{Met}(T,P) Z + s_{mix}(T,P)
\end{equation}
where $s_H(T,P)$, $s_{He}(T,P)$ and $s_{Met}(T,P)$ are the specific entropies of hydrogen, helium, and metals at the given temperature and pressure; and $X$, $Y$ and $Z$ are their mass fractions, satisfying $X+Y+Z=1$. The final term, $s_{mix}$, ensures that the total entropy recovers the ideal mixing entropy:
\begin{equation}
\label{smix}
    s_{mix}(T,P) = - \frac{k_B}{\langle A \rangle}(x\ln{x}+y\ln{y}+z\ln{z})
\end{equation}
where $x$, $y$, $z$ are the molar fractions of H, He, and metals. The mean molar weight is:
\begin{equation}
    \langle A \rangle \equiv \frac{\mu}{m_u} = A_H x+ A_{He} y + A_{Met} z
\end{equation}
where $A_H=1.0079$, $A_{He}=4.0026$ and $A_{Met}=18.0153$ are the molar masses; $\mu$ and $m_u$ are the mean molecular weight and the atomic mass constant, respectively. 

Because H/He dissolution and outgassing modify the atmospheric composition over time, the envelope abundances can deviate substantially from their initial values. This treatment contrasts with previous sub-Neptune evolution models, which typically assume a fixed envelope composition. We therefore allow $X$, $Y$ and $Z$ to vary between 0 and 1 and compute the mixture EOS across this full range. Owing to the sensitive and non-linear behavior of the mixture EOS, we adopt a high-resolution grid of $\sim$1000 points and interpolate between grid points to ensure accuracy.

In the mantle and core, the thermal structure is computed using the material's thermal expansivity, $\alpha$, following:
\begin{equation}
\label{adiabaticeq}
\left(\frac{\partial T}{\partial P} \right)_{\rm{s}}  = \frac{\alpha T}{\rho c_{\rm{P}}}
\end{equation}
where $c_{\rm{P}}$ is the specific heat capacity at constant pressure. Temperatures are initialized at the surfaces of the core and mantle, with values $T_c$ and $T_m$, and then integrated downward. Note that the derivative on the left-hand side of Eq. \ref{adiabaticeq} is staggered relative to the temperature discretization to improve convergence. Consequently, the temperature at each grid cell is solved using Brent's root-finding algorithm.

SEAMIST self-consistently evaluates solidification in both the core and mantle using their respective melting curves and solidus/liquidus. Depending on their physical state (liquid, solid, or mushy) and composition (iron, upper olivine mantle, or lower perovskite mantle), each region follows a distinct EOS. The baseline EOSs and thermodynamic parameters follow the formulation in \citet{Tang25}. 

\subsubsection{Density Reduction from Hydrogen Dissolution}
Because dissolved volatiles can significantly modify the densities of silicates and iron in liquid phases, we additionally incorporate their effects. We focus on hydrogen in this case, as we find the minimal dissolved helium mass ($<$1wt\%) usually does not impact the rock/iron core's density and thus its radius. The density of the liquid iron is evaluated by the additive volume method following a similar expression of Eq. \ref{addrho}:
\begin{equation} \label{additive}
    \frac{1}{\rho(T,P)} = \frac{f_{H,Fe}}{\rho_H(T,P)} + \frac{1-f_{H,Fe}}{\rho_{Fe}(T,P)}
\end{equation}
where $f_{H,Fe}$ is the hydrogen mass fraction dissolved in liquid iron (see Section 2.3). We evaluate the hydrogen density at each $(T,P)$ within the core using the \citet{Chabrier19} EOS. In the liquid core, we typically find $\rho_H(T,P)\gtrsim 0.8\ g/cm^3$ with a pressure $\gtrsim$1Mbar, consistent with a metallic hydrogen phase. This result aligns with our expectations and with the numerical findings in Section \ref{res:partial}, which show that hydrogen is siderophilic and highly soluble in liquid iron.

For a magma ocean, the additive-volume method of Eq. \ref{additive} is not directly applicable at low pressures. At the mantle-envelope boundary (MEB), dissolved hydrogen is expected to have a substantially higher density ($\sim0.2g/cm^{-3}$; \citealt{Hirschmann12}) than hydrogen in the gas phase ($0.1g/cm^{-3}$). This lower Gibbs free energy in the dissolved phase favors dissolution, meaning our default EOS for free $\rm{H_2}$ is not appropriate in this regime. A very recent study \citep{Stixrude25} has reported excess volumes for H$_2$-silicate mixtures under relevant pressure-temperature conditions, demonstrating that non-ideal mixing effects can be significant. Our current framework adopts a simplified density prescription to capture the leading-order structural impact of hydrogen dissolution. To achieve this, we adopt the scaling law for water-magma mixtures as a proxy. Following \citet{Bajgain15}, adding 1 wt\% $\rm{H_2O}$ decreases the melt density by 0.036$g/cm^{-3}$, nearly independent of temperature and pressure up to $\lesssim$100 GPa. Assuming that $\rm{H_2}$ dissolves predominantly as $\rm{H_2O}$, which is valid for an oxidized mantle, this implies:
\begin{equation}\label{Si-H}
    \Delta \rho_{Si-H} = -0.324g/cm^{-3} f_{H,Si}
\end{equation}
where $f_{H,Si}$ is the hydrogen mass fraction dissolved in the silicate melt.

We sanity-check this approximation using the experimental results of \citet{Hirschmann12} and the additive-volume relation. Differentiating Eq. \ref{additive} and replacing the subscript Fe with Si gives:
\begin{equation}
    \Delta \rho_{Si-H} \approx f_{H,Si}\frac{\rho_{Si}^2}{\rho_{H,Si}}
\end{equation}
Using $\rho_{H,Si}=0.2g/cm^{-3}$, $f_{H,Si}=1\%$ and $\rho_{Si}=2.5g/cm^{-3}$, we obtain $\Delta \rho_{Si-H}=0.3125g/cm^{-3}$, in good agreement with Eq. \ref{Si-H}. 

At high pressures and large dissolved hydrogen fractions, however, Eq. \ref{Si-H} can predict mixture densities lower than those obtained by assuming free-H behavior. Because the density of dissolved hydrogen should not fall below that of free hydrogen, this indicates that Eq. \ref{Si-H} underestimates the true mixture density in this regime. When this occurs, we instead apply the additive-volume method using the free-H density from \citet{Chabrier19} (similar to Eq. \ref{additive}). This approach is appropriate at high $f_{H,Si}$, where the mixture behaves more free-H-like, and at high pressures, where the metallic behavior of free hydrogen would provide a reasonable approximation to that of dissolved hydrogen.

\subsection{H/He Dissolution and Miscibility} \label{subsec:dissolution}
In each structure calculation described in Section \ref{subsec:interior}, SEAMIST self-consistently computes the dissolved H and He mass fractions, which play a crucial role in setting the planetary structure through their impact on both the envelope mass and the densities of the rock and iron layers. 

Because hydrogen solubility at the high pressures characteristic of sub-Neptune interiors remains poorly constrained experimentally, we treat the solubility as a free parameter. We explore two limiting regimes: a partially soluble case \citep{Hirschmann12,Chachan18}, in which only a finite amount of hydrogen can dissolve into the rock/iron interior and we use a multiplicative factor to scale the effective dissolution strength, and a fully miscible case \citep{Stixrude25}, in which hydrogen can be completely miscible with the rock/iron interior.

\subsubsection{Partial Soluble Models} \label{res:partial}
For the partially soluble case, we assume that hydrogen in the envelope is in chemical equilibrium with the magma ocean. As the magma ocean cools and evolves, its volatile solubility changes. When the equilibrium solubility decreases, the interior becomes supersaturated in hydrogen, and the excess dissolved volatile is efficiently mixed, transported and released into the envelope through vigorous magma-ocean convection. Because convective mixing operates on timescales much shorter than the thermal-evolution timescale, we treat this transport as effectively instantaneous and assume that the interior remains close to dissolution equilibrium throughout its evolution.

To evaluate the outgassing rates, we calculate the hydrogen solubility in the silicate melt specified by the partition coefficient $K_{\rm{D}}\equiv x_{H,Si}/x_{H,env}$, where $x_{H,Si}$ and $x_{H,env}$ are the molar fractions of molecular hydrogen in the silicate melt and the envelope, respectively. This $K_{\rm{D}}$ can be obtained from the equilibrium constant of the $\rm{H_2}$ dissolution reaction, $K_{\rm{eq}}\equiv x_{H,Si}/f_{\rm{H}}$, where $f_{\rm{H}}$ is the hydrogen fugacity. Combining the above expressions yields:
\begin{equation}\label{partition}
    x_{H,Si} = x_{H,env} f_{\rm{H}} K_{\rm{eq}}
\end{equation} 

Fugacity represents the ``effective pressure'' of a real gas and corrects for non-ideal behavior at high pressures and densities. In dissolution processes, the chemical potential of hydrogen in the gas phase is proportional to its fugacity, so $f_{\rm{H}}$ directly controls the driving force for H incorporation into the silicate melt. As a result, higher fugacity increases the equilibrium dissolved hydrogen fraction, linking the envelope thermodynamic state to mantle solubility. 

We evaluate the fugacity via the fugacity coefficient $\Phi$, defined by $f_{\rm{H}}=\Phi P$. The coefficient $\Phi$ is obtained from:
\begin{equation} \label{fugacity}
    \ln{\Phi} = \int_0^{P}\frac{Z-1}{P^\prime}dP^\prime
\end{equation} 
where $Z\equiv \mu P/(\rho k_{\rm{B}}T)$ is the compressibility factor that quantifies the deviation of the gas from ideal behavior. We use the \citet{Chabrier19} EOS to evaluate the fugacity coefficient $\Phi$ via Eq. \ref{fugacity}. Accurately determining the hydrogen fugacity requires properly characterizing the effective molecular weight of the gas. At low pressures ($\lesssim 1\ bar$), hydrogen is partially or fully thermally dissociated, with the dissociation boundary shifting to higher pressures at higher temperatures. At intermediate pressures, hydrogen remains predominantly molecular. As pressure and density increase further, reaching $\rho\gtrsim 0.1g/cm^{-3}$, hydrogen begins to undergo gradual pressure-induced dissociation. We account for these effects by incorporating the dissociation fraction from \citet{SCVH}. This treatment allows us to compute hydrogen fugacity accurately up to tens of GPa.

We then incorporate the equilibrium constant formulation derived from the experimental measurements of \citet{Hirschmann12} their Eq. 6) to compute the dissolved hydrogen mass fraction $f_{H,Si}$ via Eq. \ref{partition}. In this calculation, the experimentally derived $K_D$ begins to decline unphysically at pressures above $\sim10$GPa, a consequence of the limited pressure range ($\leq 3\,GPa$) of the underlying experiments. Following \citet{Schlichting22}, we assume that $K_D$ reaches a plateau at depth once it attains its maximum value. To account for this limitation and to explore the sensitivity of planetary evolution and radius outcomes to the strength of hydrogen solubility, we introduce a multiplicative factor $S_{sol}$ applied to $f_{H,Si}$. This parameter allows us to systematically vary the effective solubility in our models. 

To ensure a smooth and continuous behavior of the partition coefficient across the full range of pressures and temperatures, we fit the data-driven $K_D$ with an exponential function:
\begin{equation} \label{KD}
    K_D = \left[a_1\exp{\left(a_2T\right)}+a_3\right]P\exp{\left[\left(b_1/T+b_2T+b_3\right)P\right]}
\end{equation} 
where $a_1=0.0535,\ a2=-0.00088,\ a_3=0.0399,\ b_1=-65.65,\ b_2=2.364\times{10^{-6}}$ and $b_3=-0.0979$. Once $f_{H,Si}$ is determined, the total mass of hydrogen dissolved in the silicate melt is computed as $M_{H,Si} = f_{H,Si} M_{m-o}$, $M_{m-o}$ is the total mass of the magma ocean layer accounting for the dissolved volatiles.  

\subsubsection{Fully Miscible Models}
For the fully miscible case, we adopt the binodal temperature derived from the ab initio simulations of \citet{Stixrude25}. The crest of the binodal curve defines the critical temperature above which hydrogen is, in principle, fully miscible with silicate at any mixing ratio. Below this temperature, the system enters the two-phase, partial-soluble regime. We do not model this regime in our fully miscible calculations and instead examine it through our partially soluble models, for three reasons. First, our primary objective is to investigate the effects of hydrogen solubility strength rather than the detailed thermodynamic evolution along the binodal. Second, the binodal of \citet{Stixrude25} was derived for the H$_2$-MgSiO$_3$ binary system, neglecting the effects of helium, whereas our partial-solubility prescription is based on experimental measurements of hydrogen dissolution in compositionally complex, chondritic silicates. Third, as discussed in Section \ref{subsubsec:misc}, this simplification has only a minimal effect on the predicted mass-radius relationships.

Following \citet{Stixrude25}, the crest temperature is given by:
\begin{equation} \label{binodal}
    T_{b} = 4223\left(1-\frac{P}{35\,\rm{GPa}}\right) \ \ \rm{K}
\end{equation}
We neglect the coexistence of silicate in the envelope within the immiscible regime, as this occurs over a very narrow temperature interval. Relaxing this assumption provides a clear path for future extensions of the SEAMIST framework.

At each timestep, we determine the hydrogen partitioning between the envelope and the magma ocean such that the MEB temperature does not exceed the binodal temperature. This condition is enforced using Brent's algorithm, ensuring that the MEB physical conditions remains consistent with Eq. \ref{binodal}. Within each Brent iteration, we compute the hydrogen distribution between the silicate mantle and the iron core as described in Section \ref{subsubsec:partition}, which is likewise solved using Brent's algorithm.

\subsubsection{Helium Solubility and Hydrogen Partitioning between Mantle and Core} \label{subsubsec:partition}
Helium solubility in the mantle is modeled following the first-principles results of \citet{Guillot12}. Their simulations show that the dissolved helium mass fraction $f_{He,Si}$ is given approximately by
\begin{equation}
    f_{He,Si} = \frac{\gamma_{He}\rho_{Si}}{\gamma_{Si}\rho_{He}}
\end{equation}
where $\gamma_{He}/\gamma_{Si}$ is the ratio of the solubility parameters of free helium and the silicate melt. They found this ratio to be largely insensitive to pressure, allowing us to treat it as constant in our model. We also find that this approximation has a negligible effect on the overall evolution, as helium dissolution plays only a minor role (a few $0.1$ wt\%) compared to hydrogen (often $>10$wt\%). Quantitatively, the helium dissolution fractions predicted by our model are consistent with those reported by \citet{Gupta25} \footnote{where only sub-Neptunes with 1wt\% envelopes are assessed}.

The hydrogen solubility in the liquid iron is determined using the Si-Fe partition coefficient ($f_{H,Fe}/f_{H,Si}$) derived from the first-principles simulations of \citet{Luo24}. A higher core-mantle boundary (CMB) pressure increases the equilibrium dissolved hydrogen fraction in the core, leading to a larger $M_{H,Fe}$, where the dissolved hydrogen mass in the core is evaluated via $M_{H,Fe} = f_{H,Fe} M_{l-fe}$ with $M_{l-fe}$ being the total mass of the liquid iron layer. In this calculation, $f_{H,Fe}$ self-consistently scales with the prescribed solubility strength factor $S_{sol}$. 

Whenever the base of the magma ocean solidifies, we decouple volatile exchange with the iron core, since solid-state mantle convection is far less efficient at transporting dissolved gases upward than atmospheric escape is at removing them. In this regime, volatiles become effectively locked into the deep interior. On the other hand, when the mantle remains fully molten, volatiles exsolved from the iron core are efficiently transported to the envelope through vigorous magma convection. In this regime, we assume that exsolved volatile is immediately delivered into the envelope. To smoothly transition between these two regimes, we adopt a hyperbolic transition function in a form similar to Eq. 40 of \citet{Tang25}.

\subsection{Thermal and Compositional Evolution} \label{subsec:thermal}
As the interior cools, its temperature decreases, the material contracts, and the planetary radius correspondingly shrinks. This thermal evolution is modeled following the framework of \citet{Tang25}, with one key modification: in this work, we do not separately evolve the liquid and solid portions of the mantle. In the parameterized convection scheme we adopted, the temperature contrast across a thermal boundary layer between two structure layers characterizes the non-adiabatic temperature difference between them, thus characterizing the energy transport efficiency. This temperature contrast at the mantle's phase change interface (defined as a local melt fraction of 0.5) is found to be small (of order $\sim$1 K), allowing the magma ocean and the underlying solid mantle to be treated as a single adiabat. This approximation has a negligible impact on the thermal evolution and planetary radius, while yielding better numerical stability.

Because hydrogen and helium exsolve from the rock/iron interior at different rates, the envelope composition, i.e. the abundances of hydrogen, helium, and metals, evolves over time. SEAMIST tracks this coupled interior-atmosphere compositional evolution, which significantly influences the planet's radius and thermal evolution by altering the atmospheric scale height, envelope density, and cooling rate. Conversely, the evolving thermal state affects compositional evolution by modifying outgassing and solidification rates. By unifying thermal and compositional evolution, SEAMIST captures the mutual feedback between these processes.

The envelope's specific entropy, total mass, and composition determine its energy budget, comprising gravitational and thermal energy, available for cooling, making them central to thermal evolution. Its low density often makes it a major, though not exclusive, contributor to the planetary radius \citep[with the radiative atmosphere sometimes contributing equally,][]{Tang25}. We model the envelope's cooling and radius contraction via:
\begin{equation} \label{thermalenvelope}
(\frac{ds}{dt}+\dot{s}_{o} + \dot{s}_{m})\int_{M_{{\rm{ric}}}}^{M_{{\rm{ric}}}+M_{\rm{env}}}  T\,dm = - L_{\rm{RCB}} + L_{\rm{MEB}}
\end{equation}
where $M_{\rm{ric}}$ and $M_{\rm{env}}$ denote the total masses of the rock/iron core and the envelope, respectively. $T$ is the temperature of each mass shell $dm$, determined by the envelope composition and its specific entropy. On the right-hand side, we incorporate the radiative cooling at the RCB, $L_{\rm{RCB}}$, and the heating at the MEB from mantle, $L_{\rm{MEB}}$. $L_{\rm{RCB}}$ is obtained from a pre-tabulated grid of radiative-convective equilibrium models, which relates  the entropy $s$ to the intrinsic temperature $T_{\rm int}$. We extend the original grid of \citet{Tang25} with additional models to cover envelopes with up to 300$\times$ solar metallicity. As a special case, the framework is also capable of modeling water worlds using the atmospheric grid of \citet{Kempton23,Aguichine25}. Because water worlds contain little or no H/He, they are expected to be largely insensitive to the dissolution and outgassing processes investigated in this work. We nevertheless explore metallicities up to $300\times$ solar, corresponding to nearly water-dominated envelopes ($Z\sim0.8$), thereby bridging the regime between conventional H/He-rich sub-Neptunes and water-rich planets.

Compared to the pure thermal evolution framework of \citet{Tang25}, we include two additional terms, $\dot{s}_{o}$ and $\dot{s}_{m}$, to account for changes in the envelope's mixing ratio during compositional evolution. 

First, hydrogen has a higher specific entropy (per baryon) than the H-He-metal envelope mixture; outgassing of hydrogen at the MEB therefore increases the envelope entropy, while ingassing decreases it. In contrast, metals have the opposite effect because their specific entropy is lower than that of both hydrogen and helium. The effect of helium depends on the envelope's metallicity, which determines the relative specific entropy of the mixture and thus the sign and magnitude of helium-driven entropy changes. The entropy change associated with H/He outgassing is therefore:
\begin{equation}
    \dot{s}_{o} = \frac{(s-s_{H,meb})}{M_{env}}\dot{M}_{H,env} + \frac{(s-s_{He,meb})}{M_{env}}\dot{M}_{He,env}
\end{equation}
where $s$ is the envelope specific entropy, $s_{H,meb}$ and $s_{He,meb}$ are the specific entropies of hydrogen and helium at MEB conditions, $M_{env}$ is the envelope mass, and $\dot{M}_{H,env}$ and $\dot{M}_{He,env}$ are their respective outgassing rates from the rock/iron core. We find that this correction term is crucial for the envelope's energy conservation and often comparable to the entropy change from cooling.

Another contribution arises from the ideal mixing of the H-He-metal mixture. Taking the time derivative of $s_{mix}$ in Eq. \ref{smix} with respect to the evolving hydrogen and helium abundances yields a second entropy contribution in addition to $ds/dt_{o}$:
\begin{equation}
    \dot{s}_{m} \equiv \frac{ds_{mix}}{dt} = \frac{\partial{s_{mix}}}{\partial x} \frac{dx}{dt} + \frac{\partial{s_{mix}}}{\partial y} \frac{dy}{dt}
\end{equation}
The derivatives $dx/dt$ and $dy/dt$ are directly linked to the total H and He outgassing rates:
\begin{equation}
    \frac{dx}{dt} = \frac{dx}{dN_{H,env}}\frac{dN_{H,env}}{dM_{H,env}}\dot{M}_{H,env} = \frac{1-x}{M_{H,env}(1+y/x+z/x)}\dot{M}_{H,env}
\end{equation}
\begin{equation}
    \frac{dy}{dt} = \frac{1-y}{M_{He,env}(1+x/y+z/y)}\dot{M}_{He,env}
\end{equation}
The partial derivatives of $s_{mix}$ follow directly from Eq. \ref{smix}:
\begin{equation}
\begin{split}
    \frac{\partial{s_{mix}}}{\partial x} = \frac{s_{mix}}{<A>}\left(A_H-\frac{\alpha}{1+\alpha}A_{He}-\frac{1}{1+\alpha}A_{Met}\right) \\ +\frac{k_B}{<A>}\left(\ln x-\frac{\alpha}{1+\alpha}\ln y -\frac{1}{1+\alpha}\ln z\right)
\end{split}
\end{equation}
\begin{equation}
\begin{split}
    \frac{\partial{s_{mix}}}{\partial y} = \frac{s_{mix}}{<A>}\left(-\frac{\beta}{1+\beta}A_H+A_{He}-\frac{1}{1+\beta}A_{Met}\right) \\ +\frac{k_B}{<A>}\left(-\frac{\beta}{1+\beta}\ln x+\ln y -\frac{1}{1+\beta}\ln z\right)
\end{split}
\end{equation}
where $\alpha\equiv y/z$ and $\beta \equiv x/z$. We find that, in most cases, the outgassing term $\dot{s}_{o}$ dominates over the ideal-mixing term $\dot{s}_{m}$. However, the two contributions become comparable in high-solubility models when the envelope mass fraction decreases below $\lesssim 1\%$.

The hydrodynamic outflow does not directly affect the envelope's energy budget in Eq. \ref{thermalenvelope}. As shown by \citet{Tang24}, envelope convection redistributes mass and energy adiabatically, maintaining a constant specific entropy \footnote{given no cooling/heating from the top/bottom boundary is present}. Consequently, the wind primarily advects energy away without mixing material, leaving the envelope's total enthalpy per unit mass unchanged.

For the iron core and the mantle, their energy budgets are given by:
\begin{equation}
\label{ironevolution}
(c_{\rm{Pc}}\dot{T}_{\rm{c}}-\frac{\alpha_{\rm{CMB,c}} T_{\rm{CMB,c}}}{\rho_{\rm{CMB,c}}} \dot{P}_{\rm{CMB}}) \int_{0}^{M_{\rm{c}}} \frac{\partial T}{\partial T_{\rm{c}}} \bigg|_{P}\, dm = - L_{\rm{CMB}} + L_{\rm{Fe}}
\end{equation}
where $L_{Fe}$ comprises the latent heat from solidification and $L_{CMB}$ is the total energy out of the iron core. $T_{\rm{CMB,c}}$, $\alpha_{\rm{CMB,c}}$, $\rho_{\rm{CMB,c}}$ are the temperature, thermal expansivity and density of the iron core at the CMB. The derivative $\partial T/\partial T_{\rm{c}}$  is evaluated along the adiabat at the given age. Compared to Eq. 26 of \citet{Tang25}, we replace their $\dot{T}_{\rm{CMB}}$ with $\dot{T}_{\rm{c}}$, since in our framework $T_c$ tracks the CMB temperature between timesteps, whereas in their context, $T_c$ refers to a potential temperature defined at a fixed reference pressure at a given timestep.

Similarly, for the silicate mantle, we have:
\begin{equation}
\label{silicateevolution}
\begin{split}
(c_{\rm{Pm}}\dot{T}_{\rm{m}}-\frac{\alpha_{\rm{MEB,m}} T_{\rm{MEB,m}}}{\rho_{\rm{MEB,m}}} \dot{P}_{\rm{MEB}}) \int_{M_{\rm{c}}}^{M_{\rm{c}}+M_{\rm{m}}} \frac{\partial T}{\partial T_{\rm{m}}} \bigg|_{P}\, dm \\ = L_{\rm{CMB}} - L_{\rm{MEB}} + L_{\rm{Si}}
\end{split}
\end{equation}
where $L_{Si}$ includes radiogenic heating and latent heat, with their implementations described in \citet{Tang25}. In Eqs. \ref{ironevolution} and \ref{silicateevolution}, $M_c=M_{\rm{ric}}*f_{\rm{Fe}}+M_{H,Fe}$ and $M_m=M_{\rm{ric}}*(1-f_{\rm{Fe}})+M_{H,Si}+M_{He,Si}$ are the total mantle and iron masses, respectively, with $f_{\rm{fe}}$ being the iron core mass fraction and $c_{\rm{Pm}}$ and $c_{\rm{Pc}}$ are the specific heat capacities at constant pressure for the core and mantle, respectively.

We note that the second terms in the parentheses on the left-hand sides of Eqs. \ref{ironevolution} and \ref{silicateevolution} are essential when modeling volatile dissolution. As volatiles exsolve from the interior and outgas into the envelope, they increase the pressures at CMB and sometimes MEB (MEB pressure is also governed by mass loss) over time, raising the temperatures at these boundaries. This effect operates independently of the planet's secular cooling. In our calculations, we assume that the dissolved volatiles share similar thermodynamic properties with the surrounding liquid interior. The impact of this assumption is examined in Section \ref{disc:limitation}.

We model the envelope's volatile mass evolution of each species following:
\begin{equation}
\label{volatilemass}
\frac{dM_{i,env}}{dt} = -\frac{dM_{i,si}}{dt} - \frac{dM_{i,fe}}{dt} - X_{i}\dot{M}
\end{equation}
where $-dM_{i,si}/dt$ and $-dM_{i,fe}/dt$ denote the outgassing rates of species $i$ from the mantle and core, respectively, and $\dot{M}$ is the mass-loss rate driven by atmospheric escape, which is described in Section \ref{subsec:escape}. 

We solve the partially soluble models using a fifth-order ODE integrator with a stringent convergence tolerance. The variable vector includes
$M_{H,Fe}$, $M_{H,Si}$, $M_{He,Si}$, $M_{H,env}$, $M_{He,env}$, $M_{Met,env}$, $s$, $M_{\rm atm}$, $T_m$, and $T_c$, where $M_{\rm atm}$ is the mass of the radiative atmosphere. For the fully miscible models, we employ a first-order time-stepping scheme of the form $x_{i+1}=x_i+\dot{x}_i\Delta t$, which provides significantly better computational efficiency. In these models, we disable the $\dot{s}_o$ and $\dot{s}_m$ terms, as their inclusion can destabilize the numerical solution.

\subsection{Photoevaporation and Boil-Off} \label{subsec:escape}
Boil-off occurs within the first few Myr after disk dispersal. At early times, depending on the rock/iron core mass and the incident bolometric flux, the mass loss is energy-limited, either by insufficient stellar bolometric heating in the optically thin upper atmosphere or by limited interior cooling through the deep radiative layer. We model these ``bolometric-limited'' and ``intrinsic-limited'' regimes using the formulation following \citet{Tang24}. At later stages, the outflow transitions into a non-energy-limited isothermal Parker wind before eventually transitioning into photoevaporation.

The initial envelope mass fraction for boil-off is taken from the planet formation model of \citet{Ginzburg16} (GSS16 hereafter). For the partially soluble models, the initial entropy is chosen to correspond to a Kelvin-Helmholtz contraction timescale of 5 Myr. For the fully miscible models, the initial entropy is set such that it matches the entropy associated with the MEB temperature and pressure of an otherwise identical model without H/He dissolution. This results in a much longer contraction timescale ($\sim$100 Myr). Such extended cooling is expected for two reasons: (1) at fixed specific entropy, a hydrogen-poor envelope with a higher mean molecular weight contracts more slowly, and (2) significant volatile ingassing reduces the envelope mass, and a smaller envelope generally cools and contracts on a longer timescale.

SEAMIST self-consistently models the transition from boil-off to photoevaporation, which occurs once stellar XUV photons begin to penetrate down to the isothermal sonic point. SEAMIST computes photoevaporation using either full hydrodynamics at each timestep or an analytical prescription, both following the framework of \citet{Tang25b}. That work shows that any planet emerging from a boil-off phase enters a thermal-energy-mediated photoevaporation (TEMP) regime due to its low gravity, before eventually transitioning into the classic energy-limited photoevaporation phase. 

We have benchmarked the analytical prescription against the full hydrodynamic calculations within the SEAMIST framework and find that the resulting envelope mass fractions after several Gyr typically differ by only a few percent. For this reason, and for computational efficiency, we employ the analytical model in this study. Following \citet{Tang25b}, the mass-loss rate is given by
\begin{equation}
\label{mdot-scaling}
\dot{M} = 
\begin{dcases}
    \frac{\eta F_{\rm{EUV}}R_{\rm{base}}^3}{GM_p},     & \text{if } \lambda_s \geq 1\\
    \frac{\pi \alpha F_{\rm{EUV}} R_{\rm{base}}^2 \mu_s}{(\frac{\gamma}{\gamma-1}+\frac{\gamma}{2})kT_s} ,     & \text{if } \lambda_s < 1\\
\end{dcases}
\end{equation}
where $R_{\rm{base}}$ is the wind-base radius evaluated at the base pressure $P_{\rm{base}} = \mu g/\sigma_\nu$ using Eq. \ref{Rtrans}. The temperature $T_s$ and escape parameter $\lambda_s$ at the sonic point follow the analytic scalings provided in \citet{Tang25b} (T25b hereafter). The coefficients $\alpha$, $\eta$ and the characteristic photon energy used to define the wind base are also adopted from T25b. We adapt the T25b scaling for the mean molecular weight $\mu_s$ at the sonic point to account for metal-rich atmospheres:
\begin{equation} 
\label{mmw}
\mu_s = \frac{m_u}{\sum\limits_{i} X_i(1+T_s/\phi_i)/A_i} 
\end{equation}
where the index $i$ runs over hydrogen, helium, and metals. Here, $X_i$ denotes the corresponding mass fractions ($X$,$Y$,$Z$), and the coefficients are set to
$\phi_{H}=15000$, $\phi_{He}=9000$ and $\phi_{Met}=30000$. We have calibrated this relationship for Neptune-like planets, but not for very low-gravity planets where model convergence becomes challenging. Nevertheless, metal-rich envelopes are naturally associated with higher gravities, both because of their denser interiors and because their atmospheres possess a smaller scale height, which makes this scaling appropriate for the parameter space explored here.

In this work, we do not account for fractionation in the atmospheric outflow \citep{Cherubim24} or its impact on the compositional evolution of the envelope. Instead, we assume that metals, hydrogen, and helium escape at the same rate and focus on the role of volatile dissolution.

The mass-loss rates for boil-off and photoevaporation depend critically on the bolometric and XUV fluxes, respectively. For low-mass stars, bolometric luminosity evolves significantly after disk dispersal; therefore, we model stellar luminosity using the stellar evolution tracks of \citet{Spada13}. We update the irradiation temperature of the radiative atmosphere accordingly at each timestep. Our default models assume a G-type host star, whose XUV flux evolution follows the observational constraints of \citet{Ribas05}. For model suites that vary stellar mass, we compute the XUV flux using the stellar evolution model of \citet{Johnstone21}, assuming a moderately rotating star (50th-percentile rotation rate).

\subsection{Terrestrial Evolution Phase} \label{subsec:rocky}
Once the envelope becomes sufficiently thin, such that its mass is much smaller than that of the radiative atmosphere, $M_{env}<0.1M_{atm}$, either due to mass loss or envelope cooling (see Section 3.1 of \citealt{Tang25}), we transition the model to an envelope-free super-Earth regime. This criterion typically produces a smooth and continuous change in both the planetary radius and surface pressure.

After this transition, the upper boundary of the interior is set by the radiative atmosphere and stellar heating rather than the hotter envelope that carries primordial heat. In this case, a colder surface of $\sim T_{\rm eq}$ rapidly forms, which is usually solidified and thus suppresses mantle convection and thermally insulates the interior. In this regime, the atmosphere is evolved by solving for the surface partial pressures of H, He, and metals, $P_{surf,H}$, $P_{surf,He}$ and $P_{surf,Met}$. The atmospheric mass of each volatile species then follows the relation $M_{H,env}\propto P_{surf,H}$, with the proportionality coefficients computed numerically. Similar to the sub-Neptune phase, we continue to track the atmospheric mean molecular weight, interior cooling and the volatile dissolution processes throughout the super-Earth phase. SEAMIST terminates the evolution once the surface pressure drops below 1$\mu$bar. 

\section{Results} \label{sec:results}
This section is organized as follows. In Section \ref{subsec:evolution}, we describe the physics and feedback mechanisms governing the coupled evolution of thermal structure, atmospheric escape, H/He dissolution, and atmospheric composition. Section \ref{subsec:radius} quantifies how varying solubility strengths influence planetary radii across a wide range of planetary masses, stellar masses, orbital separations. In Section \ref{subsec:population}, we examine the resulting impact on the planetary occurrence-radius distribution. Section \ref{subsec:metal} examines the role of metallicity in shaping planetary radii. 

\subsection{Physics of the Coupled Evolution} \label{subsec:evolution}
\subsubsection{Partially Soluble Case}
\begin{figure*}
\centering
\includegraphics[width=0.9\textwidth]{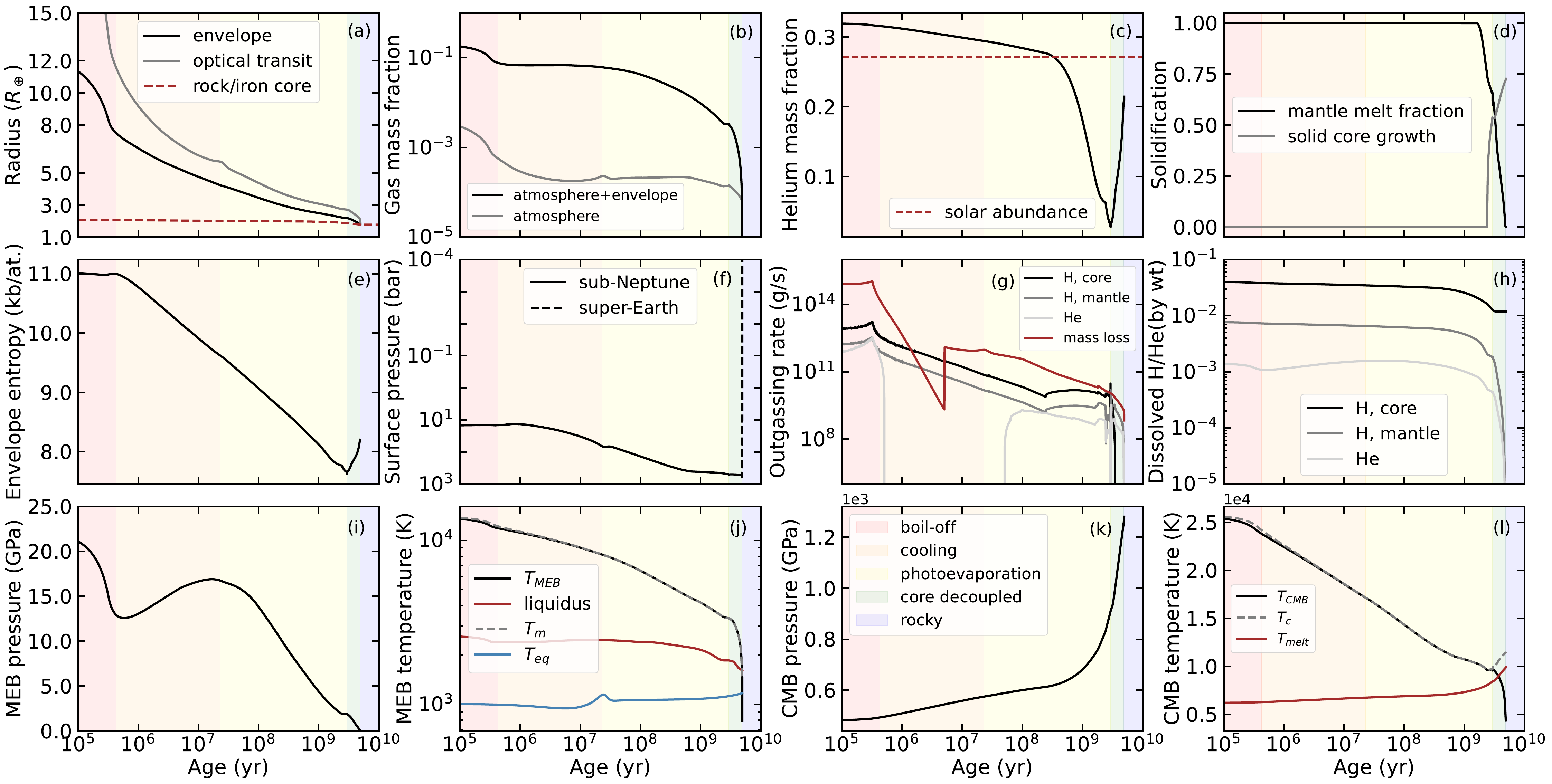} 
\caption{ Example evolution tracks of a model planet with a core mass of $8.8M_\oplus$, an incident flux of $300F_\oplus$, and a solubility strength three times the default value. Panel (a) shows the evolution of the envelope radius (black), optical transit radius (gray), and rock/iron core radius (red). Panel (b): the total gas mass fraction (envelope + radiative atmosphere; black) and the mass of the radiative atmosphere (gray). Panel (c): the helium mass fraction. Panel (d): the mantle bulk melt fraction (black) and the inner solid core radius normalized to the total core radius (gray). Panel (e): the envelope specific entropy. Panel (f): the surface pressure, defined at the RCB for sub-Neptunes and at the mantle surface for envelope-free super-Earths \footnote{during the transition between these regimes, the RCB converges to the MEB}. Panel (g): the H outgassing rates from the iron core (black) and mantle (gray), the He outgassing rate from the mantle (light gray), and the atmospheric mass-loss rate (red). Panel (h): the corresponding dissolved mass fractions in the core and mantle. Panel (i): the pressure at the MEB. Panel (j): the MEB temperature (black), the mantle potential temperature (gray dashed), and the mantle liquidus (red). Panel (k): the CMB pressure. Panel (l): the CMB temperature (black), the core potential temperature (gray dashed), and the core melting curve (red). Depending on the dominant physics, we divide the evolution into five phases, indicated by shaded regions: boil-off-dominated (red), cooling-dominated (orange), photoevaporation-dominated (yellow), core-decoupled (green), and super-Earth (blue). See the main text for discussion.
}
\label{evo_coup}
\end{figure*}

In Figure \ref{evo_coup}, we present evolution tracks for an $8.8\,M_\oplus$ planet with a solar-metallicity envelope, receiving an insolation of $300\,F_\oplus$ from a Sun-like star. This example illustrates the coupled effects of atmospheric escape, thermal evolution, volatile dissolution and outgassing, compositional evolution, and interior solidification within our framework. Based on these processes, we identify five distinct evolutionary phases. Note that the characteristic timescales and relative importance of these phases depend sensitively on planetary properties, including mass, incident flux, composition, and host-star type. Consequently, not all planets experience all five phases, and the timing of each phase can vary substantially across the model population. We selected this particular case because it exhibits all five phases and therefore provides a useful illustration of the diverse evolutionary pathways captured by our framework, rather than a universal evolutionary sequence applicable to all sub-Neptunes.

The interplay among thermal evolution, atmospheric escape, compositional evolution, and H/He outgassing is inherently complex and cannot be captured by simple analytic arguments generalized for all planets; a definitive description requires fully self-consistent numerical simulations. Here, we qualitatively highlight several feedback mechanisms that commonly emerge in our model planets.

We find that the MEB pressure serves as a useful diagnostic for identifying the evolutionary phase of a planet. The MEB pressure is highly sensitive to the envelope mass: mass loss reduces the envelope's weight, thereby lowering the pressure exerted on the magma ocean. In contrast, planetary cooling and the associated thermal contraction increase the mean envelope density, which gradually raises the MEB pressure over time. The competition between these effects governs the MEB evolution and helps delineate the dominant physical regime at any given epoch.

The red shaded region in Figure \ref{evo_coup} marks the boil-off phase (usually lasting for $\lesssim$3 Myr). During this stage, both the planetary radius (panel a) and the gas mass fraction (panel b) rapidly decrease due to extremely efficient mass loss. Consequently, the MEB pressure (panel i) drops sharply. For this example planet, the outgassing rate is relatively weak compared to the escape rate and therefore does not significantly influence the evolution at early times, even though outgassing is generally stronger compared to later ages. In other planets, however, outgassing can exceed the mass-loss rate and trigger a catastrophic boil-off event (see Section \ref{subsubsec:bf}). The specific entropy (panel e) changes only modestly during this phase because the mass-loss timescale is much shorter than the thermal evolution timescale, resulting in a temporary decoupling between the two processes (see \citealt{Tang24}). In more highly irradiated and lower-mass planets than the one shown here, the entropy may exhibit a brief increase, which can further enhance mass-loss efficiency. The helium abundance (panel c) begins at a super-solar value due to the preferential dissolution of hydrogen relative to helium into the magma ocean. Meanwhile, the MEB temperature decreases, despite the nearly constant entropy, because the mantle experiences rapid depressurization as the envelope mass declines.

The second phase (orange) is a cooling-dominated stage characterized by minimal total mass loss, since the photoevaporative timescale is typically longer than tens of Myr. This behavior is reflected in panels b and e. As the planet cools, thermal contraction increases the MEB pressure (panel i), but the declining MEB temperature (panel j), driven by envelope cooling, still allows interior hydrogen to slowly exsolve, thereby reducing the helium mass fraction (panel c). During and after this phase, the CMB pressure (panel k) steadily increases as the rock/iron interior continues to contract. Meanwhile, the RCB location (panel f) shifts progressively deeper as the envelope cools. Note that in less-irradiated and high-mass planets, which are not susceptible to photoevaporation, solidification may occur in this phase.

The evolution can then proceed into a photoevaporation-dominated phase (yellow), during which the MEB pressure (panel i) begins to decline again as mass loss becomes efficient. As the envelope thins, the escape rate gradually decreases, but the outgassing rate does not decline at the same pace. This is because depressurization of the magma ocean rapidly exsolves hydrogen from the rock/iron interior (panel g). Cooling further enhances this process through progressive mantle solidification, as hydrogen is essentially insoluble in solid silicates and is therefore released immediately upon freezing. Consequently, the helium abundance (panel c) begins to drop rapidly. Mantle solidification (panel d) is typically triggered once the envelope becomes thin enough that it no longer provides strong thermal insulation. A solidified mantle bottom generally delays and slows subsequent core solidification (see \citet{Tang25}), although in this example (Fig. \ref{evo_coup}) the onset of iron solidification happens to coincide with mantle freezing. When the outgassing rate becomes comparable to the atmospheric escape rate, the total gas mass fraction exhibits a brief plateau, typically at an envelope mass fraction of $\sim$ 0.3\%. The rapid exsolution of hydrogen from the deep interior increases the volatile content in overlying layers, which steepens the rise of the CMB pressure. As a result, we frequently observe an accompanying increase in the core potential temperature $T_c$ (see Eq.\ref{silicateevolution} and the surrounding discussion).

When the bottom of the mantle freezes, cooling and volatile outgassing of the iron core greatly slows down due to inefficient mantle convection at the CMB, ultimately cutting off the core's mass and energy supplies to the envelope. At this point, the core becomes thermally and chemically decoupled from the planet's subsequent evolution (green phase). This stage typically begins when the gas mass fraction has fallen to a few times $0.1\%$. Without the core's heat flux, the mantle cools rapidly (panel j) leading to a large temperature contrast across the MEB, as seen in $T_c$ (gray dashed) and $T_{CMB}$ (black) of panel l. This cooling can dominate over the envelope's own gravitational contraction, causing the envelope to heat up and its specific entropy to increase (panel d). The combination of this thermal inflation and the cessation of volatile replenishment from the core (panel g) strongly promotes the final transition to the super-Earth regime. This transition proceeds more slowly in weakly irradiated planets. In addition, once core-envelope hydrogen exchange is cut off, hydrogen and helium exsolve from the magma ocean at more comparable rates (panel g), causing the helium mass fraction to rise again (panel c). The mantle generally finishes solidifying before the planet completes its transition to a super-Earth, which occurs once the gas mass fraction drops to a few times $ 0.01\%$.

The final transition into the super-Earth regime typically occurs once the total gas mass fraction falls in the range of 0.001-0.01\%, though lower-mass planets can undergo this transition at somewhat higher gas fractions. During this stage, the planetary radius reduces by an additional 20-30\% through hydrodynamic outflow, approaching the bare rock/iron core radius. We generally find that the super-Earth phase is much shorter than the preceding envelope-bearing stage. In the example model, this phase lasts only $\sim 70$ Myr. However, for very low-mass and weakly irradiated planets, such as Earth analogues, this phase may persist for up to gigayears, making it comparable in duration to the sub-Neptune epoch. Throughout this stage, the MEB temperature (known as the surface temperature in this phase, panel j) gradually cools toward the planet's skin temperature ($<T_{\rm eq}$) by the end of the evolution. The equilibrium temperature $T_{\rm eq}$ is shown in blue. The bump at $\sim$10 Myr coincidentally coincides with the phase transition and reflects the onset of sustained stellar nuclear burning following the pre-main-sequence contraction.

\subsubsection{Catastrophic Boil-Off} \label{subsubsec:bf}
We identify a strong positive feedback loop between boil-off and outgassing that operates on timescales far shorter than those of standard boil-off evolution. This runaway behavior typically emerges when the solubility is high.

\begin{figure*}
\centering
\includegraphics[width=0.9\textwidth]{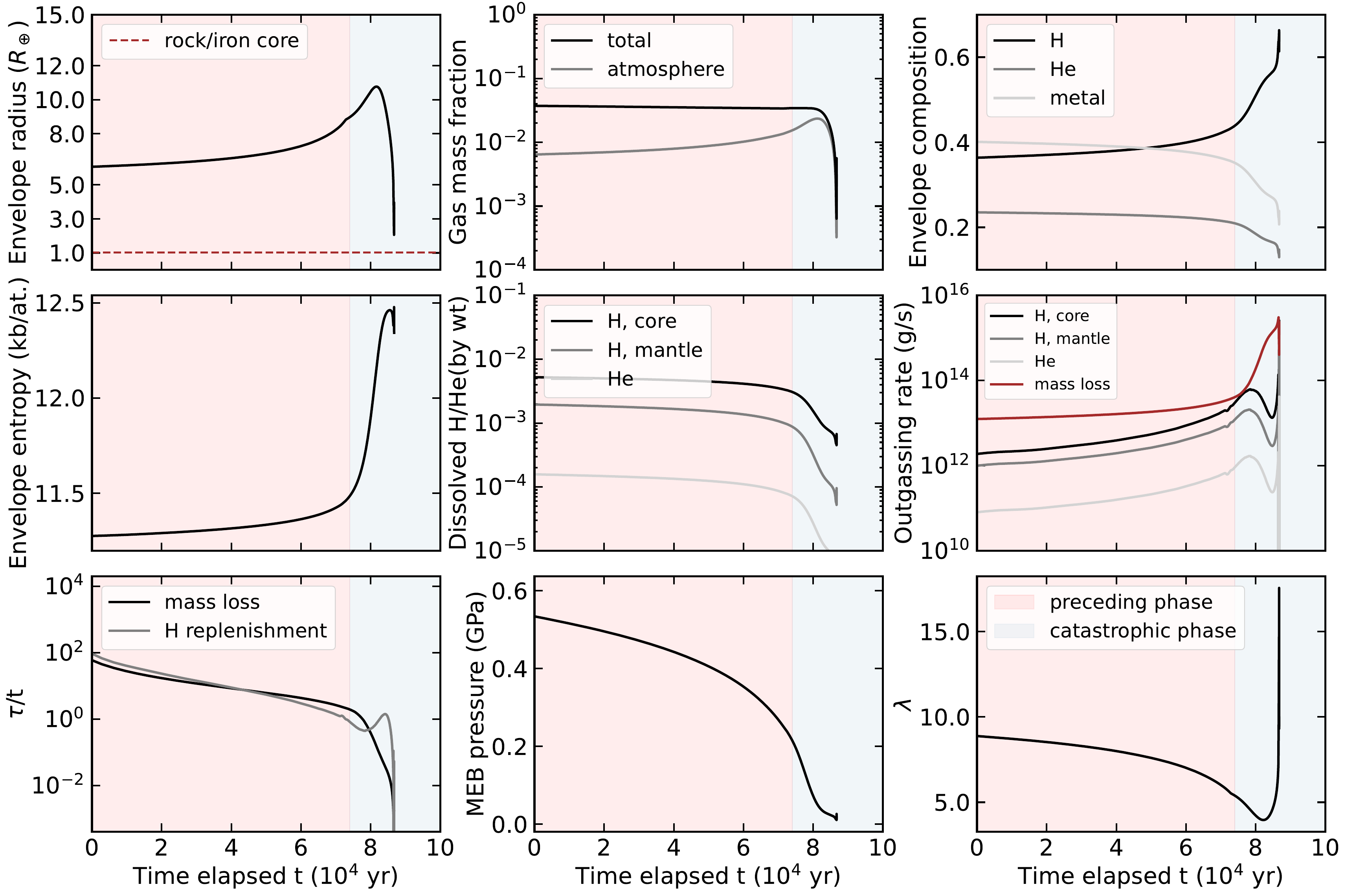} 
\caption{ Example evolution tracks of a planet undergoing catastrophic boil-off, triggered by a positive feedback between an isothermal Parker wind and hydrogen outgassing. The x-axis shows the time elapsed since $10^5$ yr after disk dispersal. Panels display the envelope radius (top left), gas mass fractions (top middle), atmospheric composition by mass fraction (top right), envelope specific entropy (middle left), dissolved volatile mass fractions (middle), outgassing and mass-loss rates (middle right), mass-loss and hydrogen-replenishment timescales normalized to the elapsed time (bottom left), MEB pressure (bottom middle), and the atmospheric escape parameter (bottom right). We divide this evolutionary phenomenon into two phases, marked by shaded regions: an initial phase (driven by either boil-off or photoevaporation; red) characterized by comparable outgassing rate and escape rate and, once the hydrogen replenishment timescale becomes shorter than the elapsed time, a catastrophic boil-off phase (blue). See the main text for discussion.
}
\label{evo_bf}
\end{figure*}

Figure \ref{evo_bf} illustrates an example: an Earth-mass planet with a 30 times solar-metallicity envelope irradiated at 30 $F_\oplus$. In this case, the solubility is set to 10 times our default value. The x-axis indicates the time elapsed since $10^5$ yr after the onset of disk dispersal, corresponding to the moment when the disk has fully dispersed (assuming a disk dispersal timescale of $10^5$ yr).

We divide the boil-off evolution into two distinct phases. The early phase (red) corresponds to a period in which ``pressure'' gradually builds until mass loss becomes catastrophic. During this stage, boil-off removes envelope mass on a characteristic timescale of $10^5$ yr, while volatiles exsolve at a comparable rate. Consequently, the envelope mass fraction decreases only modestly, by about 10\% (top middle). Outgassing increases the envelope's specific entropy (middle left) and reduces both its mean molecular weight and density because of the rising H abundance (top right). This abundance change reflects the premature loss of He and metals. These changes enlarge the atmospheric scale height, inflating the envelope radius (top left) and thereby accelerating mass loss (bottom right) through an increase in the density at the sonic point, $\rho_s$. This enhancement is exponential: hydrostatic equilibrium gives the scaling $\rho_s\sim\rho_{\rm RCB} \exp{[-(R_s-R_{\rm RCB})/H]}$. As the mass loss rate rises (middle right), the MEB pressure (bottom middle) falls further due to mass removal, which in turn accelerates outgassing (middle right). Outgassing also feeds back on itself by lowering the interior density, contributing to the continued decrease in MEB pressure. Eventually, the replenishment of H from outgassing overtakes the hydrogen escape (bottom left), causing the positive feedback to strengthen over time.

When hydrogen replenishment operates on a timescale comparable to the time elapsed (bottom left), boil-off enters a catastrophic phase: the planet is effectively doomed to rapidly decay into a super-Earth. Early in this phase, the mass-loss rate remains comparable to the outgassing rate (middle right), leading to only a slowly evolving envelope mass fraction (top middle). However, this quasi-equilibrium cannot be maintained once the interior hydrogen reservoir begins to deplete. The balance then collapses, the stored ``pressure'' is released, and the envelope lifetime shortens rapidly, reaching 1 Kyr. This stage is marked by a growing contrast between the rising mass-loss rate and the falling outgassing rate (top right). The mass-loss rate increases despite the shrinking planetary radius because the mean molecular weight continues to decline: helium and metals escape more efficiently while their supply from the interior is limited. The catastrophic blow-off ends only after the interior hydrogen has been sufficiently depleted that the hydrogen abundance, and thus the mass-loss rate, begins to drop, forming a temporary quasi-equilibrium between outgassing and mass loss. This balance is typically unstable and gives rise to oscillations in the gas mass fraction. The system ultimately transitions to a super-Earth at a remarkably small remaining envelope mass fraction of a few $0.1\%$. 

In summary, in the absence of solubility, a planet would undergo a normal, moderately strong boil-off. However, hydrogen dissolution initially increases the envelope's mean molecular weight and introduces a compensating outgassing flux, which together slow and delay the onset of mass loss. As the planet evolves and conditions gradually shift to favor mass loss, this delicate balance eventually collapses. Once broken, the stored ``pressure'' is released on a much shorter timescale than in a standard boil-off, driving the system into catastrophic phase. 

Consequently, we find that this catastrophic phase is more common in high-metallicity planets, systems in which a boil-off can only be initiated if the core mass is sufficiently low or the incident flux is sufficiently strong. Strong solubility is also required to trigger this instability: it raises the initial mean molecular weight, suppressing early boil-off, and later buffers the envelope in a way that enables the delayed runaway. Increasing solubility strength generally pushes the onset of the catastrophic phase to later evolutionary times. In the fully miscible models, this catastrophic transition commonly appears at late ages in low-mass planets ($\lesssim 3M_\oplus$), naturally leading to their rapid evolution into super-Earths.

\subsubsection{Fully Miscible Case} \label{subsubsec:misc}
\begin{figure}
\centering
\includegraphics[width=0.5\textwidth]{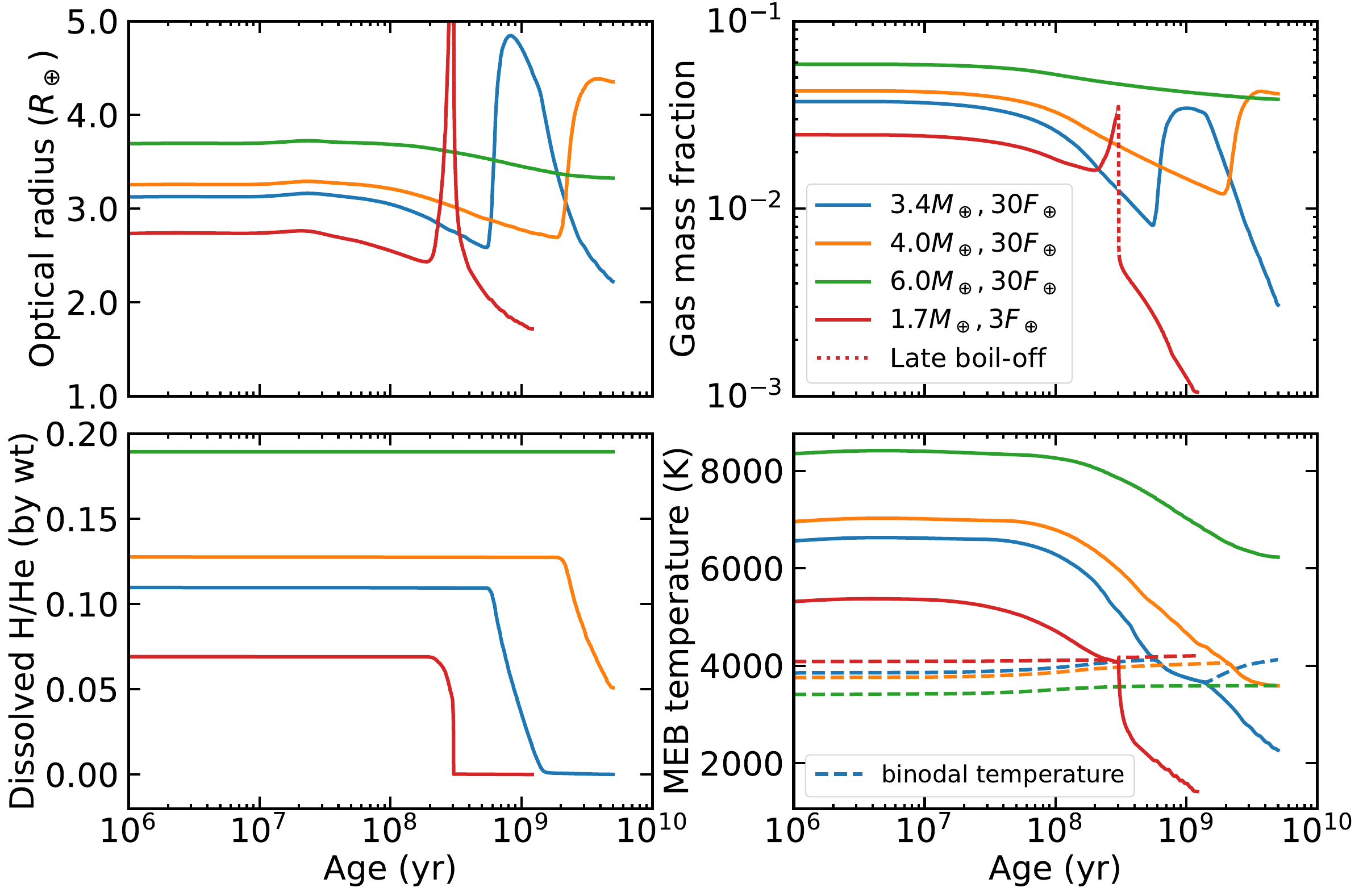} 
\caption{ Example evolution tracks for fully miscible models. We identify three distinct evolutionary regimes and show representative planets for each. All planets are irradiated at the same level of 30$F_\oplus$ but have different core masses: a cooling-dominated case (green, 6$M_\oplus$), an outgassing-dominated case (orange, 4$M_\oplus$), and an escape-dominated case (blue, 3.4$M_\oplus$). We also show a lower-mass escape-dominated planet (1.7$M_\oplus$, also irradiated with a lower 3$F_\oplus$), in which a late catastrophic boil-off is triggered and accounts for the majority of the total mass loss. Panels show optical radius (top left), gas mass fraction (top right), total dissolved volatile mass fraction (bottom left), and MEB temperature (solid) and binodal temperature (dashed) in the bottom right. See the main text for discussion.
}
\label{evo_misc}
\end{figure}

The evolution of fully miscible planets differs markedly from that of partially soluble planets. We show several example evolution tracks in Figure \ref{evo_misc}. In general, we find that the rock/iron core radius is inflated by 50-100\% due to the presence of dissolved hydrogen. In contrast, the density reduction in partially soluble models with $S_{\rm sol}\leq3$ is usually small ($\lesssim$3\%), which correspond to a radius reduction of $\lesssim$1\%.

Unlike the model planets in \citet{Rogers25}, which assume a pure-hydrogen envelope resulting in an envelope mass fraction of $<1\%$ after dissolution, our model predicts substantially more massive envelopes (5-10\% by weight) due to the presence of undissolved He and metals. These species provide thermal insulation, maintaining a hot mantle that keeps essentially all interior hydrogen dissolved. This thermal insulation significantly delays hydrogen outgassing compared to previous models. Consequently, the MEB temperature remains well above the hydrogen binodal temperature during this phase. The envelope is therefore initially dominated by helium and, at sufficiently high metallicity, may also contain a substantial metal component. For planets with intermediate or high core masses, the evolution is primarily cooling-dominated. For example, a $6M_\oplus$ planet (green curve) retains hydrogen tightly locked in its rock/iron core throughout the evolution (bottom left), due to the large difference between the binodal temperature and the MEB temperature (bottom right). Atmospheric escape is generally inefficient (top right), as hydrogen ingassing reduces both the planetary radius (top left) and the atmospheric scale height through the increased mean molecular weight.

However, with a core mass decreased to $4M_\oplus$ (orange), cooling accelerates, and hydrogen begins to outgas rapidly at a late age. This outgassing not only causes an abrupt increase in the envelope mass (top right) in the absence of efficient mass loss, but also lowers the mean molecular weight, leading to a rapid expansion of the planetary radius (top left). This radius inflation and sudden outgassing contrasts with the pure-hydrogen case of \citet{Rogers25}, as the onset of outgassing in our models is primarily controlled by the helium content in the envelope rather than a continuous self-regulation. Once a new equilibrium at higher envelope mass is established, the planet evolves smoothly, with the MEB temperature closely tracking the hydrogen binodal temperature. This behavior indicates the coexistence of hydrogen in both the mantle and the envelope, while the envelope transitions from helium-dominated to hydrogen-dominated. In this case, we call the evolution outgassing-dominated.

As the core mass is further decreased to $3.4M_\oplus$ (blue), the radius inflation caused by hydrogen outgassing triggers efficient photoevaporation. In this case, the hydrodynamic outflow rapidly shrinks the planetary radius, which we call escape-dominated evolution. The envelope mass loss accelerates interior cooling, and together with the depletion of dissolved hydrogen, the MEB temperature drops below the binodal temperature. In this case, hydrogen is expected to become partially soluble at a small mixing ratio, a regime that is not captured by our fully miscible models. We find, however, that this transition occurs only after the envelope mass fraction has declined to below $\sim1\%$, at which point the mass-loss timescale becomes sufficiently short that the planet rapidly loses its remaining hydrogen envelope and evolves into a super-Earth. In mass-radius space, this transition therefore occupies only a narrow range of planetary masses ($<1M_\oplus$) in Figure \ref{MR}. Consequently, neglecting the partially soluble regime in our fully miscible models has a negligible impact on the radius planes presented in Section \ref{subsec:radius}.

For even lower core masses, such as $1.7M_\oplus$, the radius inflation from hydrogen outgassing triggers a catastrophic boil-off, rapidly removing the envelope down to a few 0.1\% by weight, after which photoevaporation removes the remaining gas. 

\subsection{Impact of Solubility Strength on Planetary Radius Across Parameter Space} \label{subsec:radius}
Three key physical parameters, the rock/iron core mass $M_{\rm ric}$, bolometric flux (or orbital separation) $F$, and stellar mass $M_s$, govern the formation of the radius gap, as they critically determine the strength of atmospheric mass loss. In this section, we run about 1000 model planets covering a wide region in parameter space, to show the influence of those parameters on radius.

\begin{figure*}
\centering
\includegraphics[width=0.9\textwidth]{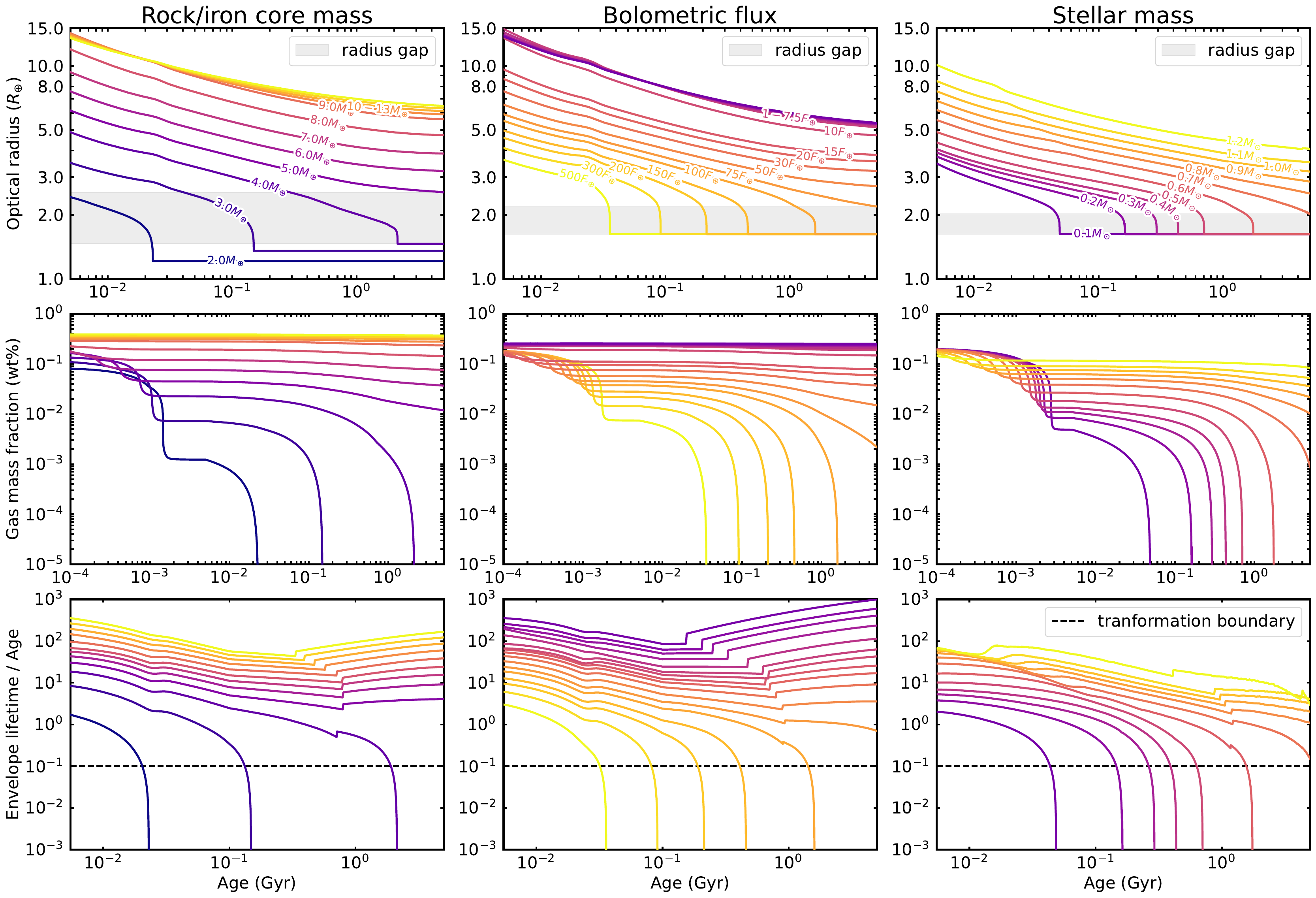} 
\caption{  We present suites of evolutionary tracks in each column to illustrate the formation of radius gap in high-dimension space. The top row shows the optical transit radius, the middle row shows the total gas mass fraction, and the bottom row shows the envelope lifetime normalized by the planetary age. In the left column, we vary the rock/iron core mass linearly over the range 2-13$M_\oplus$, while fixing the host star to be Sun-like and the irradiation to $30F_\oplus$. In the middle column, we assume a Sun-like star and a core mass of 6$M_\oplus$, and vary the bolometric flux over the range 1-500$F_\oplus$ in a quasi-logarithmic manner. In the right column, we vary the stellar mass linearly over the range 0.1-1.2$M_\odot$, while assuming an irradiation of $30F_\oplus$ and a core mass of 6$M_\oplus$. In each column, the radius evolution exhibits a clear dichotomy: planets either follow a convex evolutionary track and retain substantial envelopes, ending as sub-Neptunes, or follow concave tracks that correspond to rapid envelope loss and a final super-Earth outcome. The resulting radius gaps are indicated by the shaded regions. This example corresponds to the no-solubility case, and similar behavior is found in partially soluble models. All models have solar metalicities.
}
\label{tracks}
\end{figure*}

Figure \ref{tracks} shows evolution tracks in which we vary $M_{\rm ric}$, $F$, and $M_s$ in the left, middle, and right panels, respectively. For the bolometric-flux case, we fix the stellar type to a Sun-like star; this is effectively equivalent to varying the orbital separation. The top, middle, and bottom panels show the evolution of the optical transit radius, gas mass fraction, and the ratio of envelope lifetime to planetary age.

The origin of the radius gap arises from a dichotomy in radius evolution: planets that are less susceptible to mass loss follow a smooth, gradually slowing contraction, whereas more mass-loss-susceptible planets undergo a rapid radius decline once their envelopes become sufficiently thin. The latter case occurs once envelope lifetime becomes shorter than the planetary age, as seen in the bottom panels where the y-axis is the envelope lifetime normalized by the age. A planet is inevitably transformed into a super-Earth through mass loss when the lifetime-age ratio approaches 0.1 (black dashed line), leading to a nearly vertical drop in radius (top) and envelope mass (middle). In the top panel, these two evolutionary behaviors are separated by the gray shaded region, which marks the location of the radius gap. The observed radius gap emerges from the combined influence of all three parameters and their underlying distributions.

We then examine how the radius responds to each parameter individually in Figures \ref{MR}, \ref{PR} and \ref{sMR}. To do this, we compute a finer grid of evolution models, extract the planetary radius at 5 Gyr for each track, and present the resulting radius surfaces across the parameter space. We similarly compute the corresponding gas mass fraction surface. These radius surfaces represent theoretical upper limits. Any planet lying above a given surface is strictly forbidden, as its radius and gas mass fraction would have been substantially reduced by mass loss. In contrast, a planet located below a surface remains viable, it could exist if it formed with a gas mass fraction lower than the theoretical upper limit predicted by our model, which is no more than the mass fraction predicted by planet formation.

\subsubsection{Planetary Mass} \label{subsec:mass}
\begin{figure*}
\centering
\includegraphics[width=0.9\textwidth]{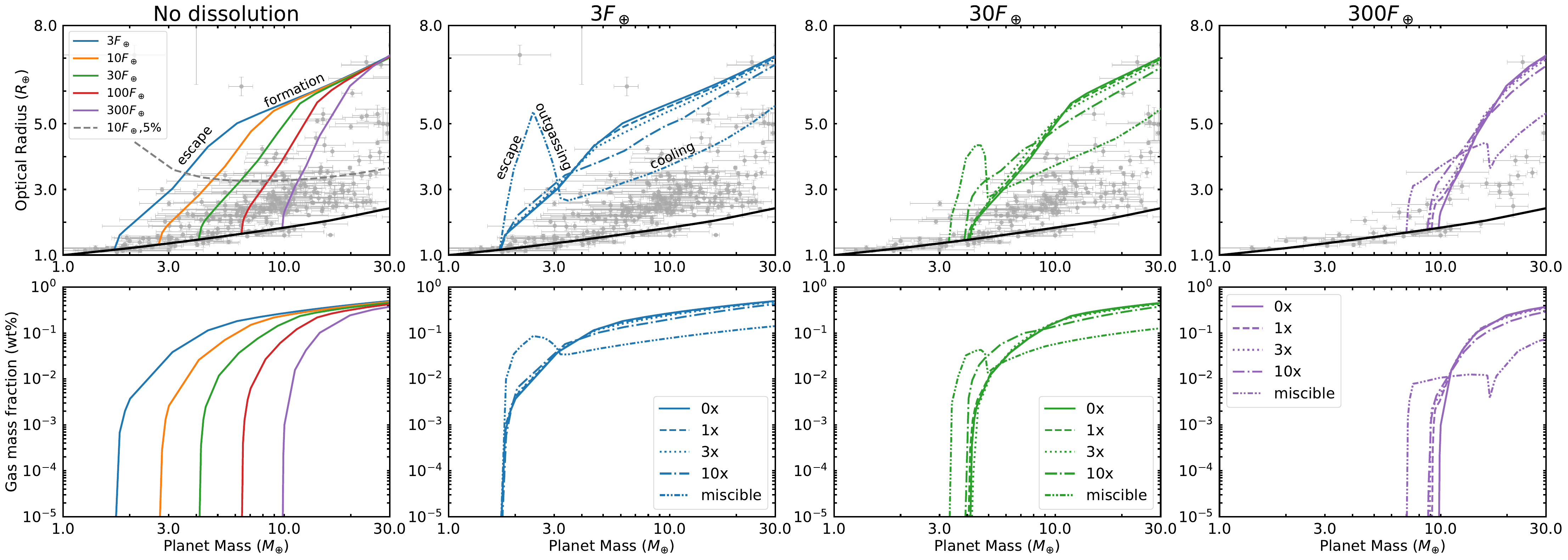} 
\caption{ Mass-radius (M-R) relationships (top row) evaluated self-consistently after 5 Gyr of coupled evolution, along with the corresponding gas mass fractions (bottom row). The left panels show M-R curves with no dissolution for varied incident fluxes. Each curve exhibits an escape-dominated segment at low masses and a formation-dominated segment at higher masses. A similar behavior is seen in the right panels as well. The background data include all confirmed planets with well-measured masses. A traditional M-R curve with a fixed envelope mass fraction of 5\% (black dashed) is shown for comparison. The radius of a bare rock/iron super-Earth with Earth-like composition is shown as a black solid line. In the right three panels, the bolometric flux is fixed, and the solubility strength is varied: no dissolution (solid), default solubility (dashed), 3x default (dotted), 10x default (dash-dot), and fully miscible (dash-double-dot). The background data are filtered to match the corresponding flux level. For the fully miscible models, the three regimes (cooling-dominated, outgassing-dominated, and escape-dominated) corresponding to the example evolutions in Figure \ref{evo_misc}, are labeled. All models have solar metalicities. See main text for discussion.
}
\label{MR}
\end{figure*}

Figure \ref{MR} shows the resulting mass-radius (M-R) relationship (top panels) and corresponding envelope mass fractions (bottom) across a range of incident bolometric fluxes. Atmospheric escape becomes stronger at lower $M_{\rm ric}$ because the planetary potential well is shallower, facilitating mass loss, and because the larger atmospheric scale height produces a more extended photoionization base.

In the leftmost panels, we compare models with varying flux while assuming no solubility. In the right three panels, we fix the flux in each column and additionally vary the solubility strength. In the background, we plot all confirmed exoplanets with well-measured masses. The confirmed planets table is sourced from \citet{NasaExoplanetArchive}. For the right panels, we include only the observed planets receiving a flux higher than the corresponding model flux. At low masses, planets transform into super-Earths, with their radii set by the rock/iron core (black curves, assuming no dissolved volatiles). This transformation boundary shifts to higher masses as the incident flux increases.

The observed distribution aligns well with our results: very few planets lie above our radius surfaces in any panel, and the slopes of the data are consistent with our predictions. For comparison, a classical M-R curve (gray dashed), evaluated at a fixed composition, fails to reproduce this positive slope because it does not encode any information about interior composition. 

Each radius curve consists of two segments: a steeper segment at the low-mass end, set by atmospheric escape, and a shallower segment at the high-mass end, which is largely insensitive to escape and instead reflects the initial mass fractions predicted by formation models. The escape-dominated segment matches the observations particularly well, whereas the formation-dominated segment does not reproduce the overdensity of planets at smaller radii well. This mismatch appears to be more pronounced in the high-irradiation case (rightmost panels). We discuss the possible reasons in the Discussion.

At intermediate to low solubility strengths (dashed and dotted curves), the M-R curves shift only slightly from the no-dissolution models (solid curves). This contrasts with previous expectations: outgassing volatiles are thought to buffer mass loss, prolonging the envelope lifetime \citep{Chachan18} and lowering the core-mass threshold for transition. Furthermore, planets with volatiles stored in the rock/iron core were expected to experience less vigorous escape due to the reduced envelope radius. In these models, the dissolved hydrogen fraction typically ranges from 1-10\% by weight in the iron core ($f_{H,Fe}$) and is about an order of magnitude smaller in the mantle, sufficient to affect the radius.

The apparent discrepancy arises from the new feedback mechanism identified in Section \ref{subsec:evolution}: as the envelope becomes increasingly hydrogen-rich due to outgassing, the mass-loss rate accelerates. At intermediate to low solubility strengths, this new effect balances the previously proposed buffering effect and the reduced radius by volatile ingassing, resulting in M-R curves that remain similar to the no-dissolution case.

At higher solubility strengths (10 times the default, dash-dot curves), the M-R curves shift more noticeably, though the overall change remains modest. These models correspond to a core hydrogen mass fraction $f_{H,Fe}$ of $\sim$20\%. With increasing solubility, the formation-dominated segment of each curve shifts to lower radii, while the escape-dominated segment rises, slightly extending the super-Earth transformation boundary to lower planetary masses. This extension is particularly pronounced for planets experiencing high incident flux.

For the fully miscible case (dash-double-dot curves), a substantially larger fraction of hydrogen is dissolved in the mantle relative to the core compared to the partially soluble models. We divide each radius curve into three segments: the cooling-, outgassing-, and escape-dominated regimes, following the discussion in Section \ref{subsubsec:misc}. In the cooling-dominated regime, complete hydrogen dissolution produces a significantly smaller radius, typically below $5R_\oplus$, compared to the other curves. Outgassing generates a peak in the lower-mass portion of the curve: the rising (right) side of the peak is driven by outgassing, while the dipping (left) side results from escape. This peak is more pronounced in less-irradiated planets, where strong escape maintains a similar super-Earth transformation boundary as in other models. In highly irradiated planets (right panels), the boundary extension is more pronounced, and the peak is correspondingly weaker. However, no observed planets appear within the peak. We return to this in the Discussion.

The sharp cutoff defining the left boundary of the peak corresponds to the escape-dominated phase. During this phase, when the hydrogen inventory becomes sufficiently small for the MEB temperature to fall below the crest of the binodal, H dissolution formally enters the partially soluble regime. As discussed in Section 3.1.3, we do not model this regime explicitly because it has little impact on the resulting M-R relationships, as indicated by the similar slopes of the fully miscible and partially soluble models in Figure \ref{MR}.

\subsubsection{Orbital Period and Flux}
\begin{figure*}
\centering
\includegraphics[width=0.9\textwidth]{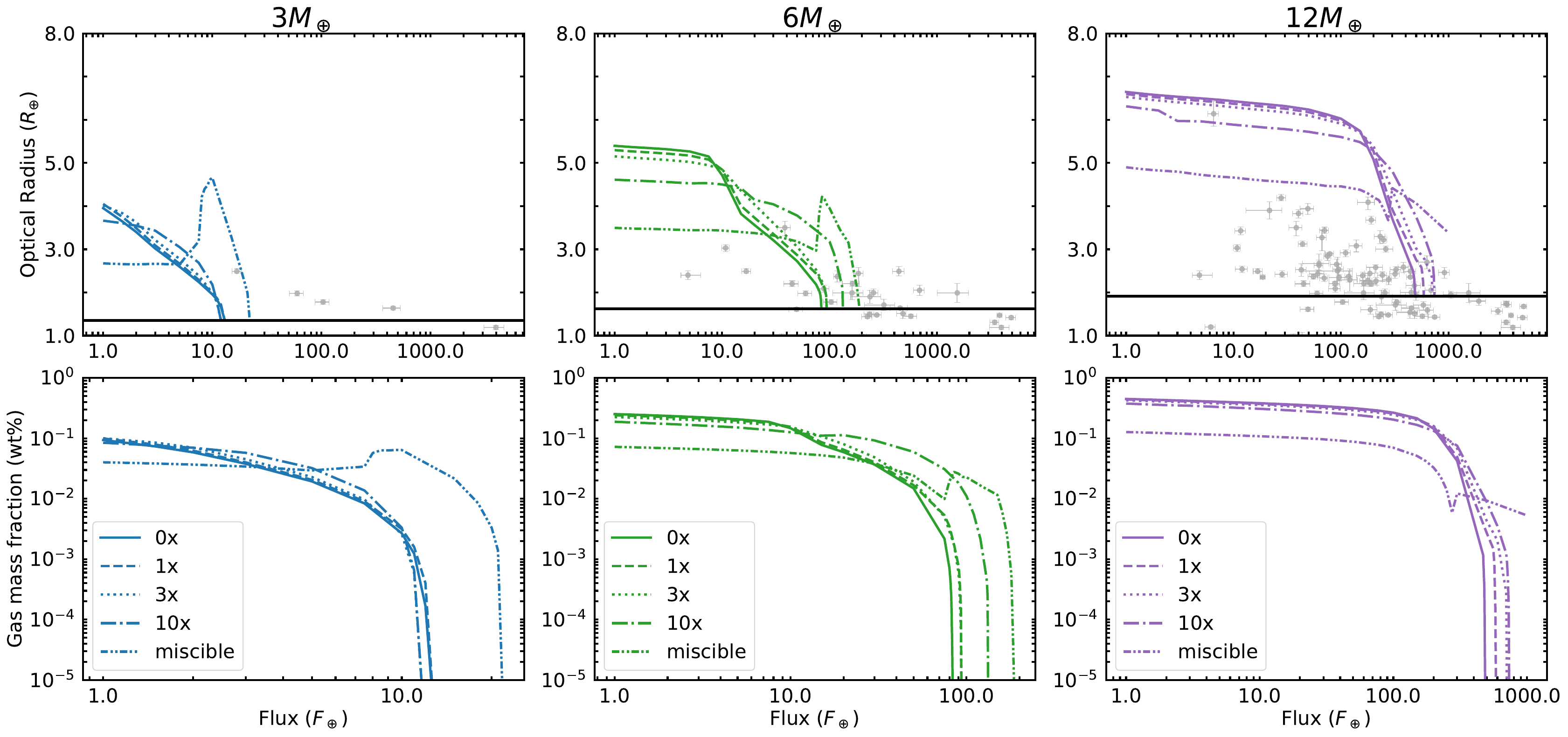} 
\caption{ Flux-radius (F-R) relationships (top row) and the corresponding gas mass fractions (bottom row) for planets orbiting sun-like stars. The setup is analogous to that in Figure \ref{MR} for the M-R relationships. Each panel shows flux-R curves for low-mass (3$M_\oplus$, left), intermediate-mass (6$M_\oplus$, middle), and high-mass (12$M_\oplus$, right) cores. For partially soluble models, the flat segments correspond to formation-dominated planets, whereas the sloped segments at shorter periods are escape-dominated. In each panel, we also plot all confirmed exoplanets with measured masses below the corresponding core mass shown in that panel for comparison. All models have solar metalicities. See main text for discussion.
}
\label{PR}
\end{figure*}

Figure \ref{PR} illustrates how incident bolometric flux and orbital period influence planetary radius. Because we assume a Sun-like host star, the orbital period directly reflects the bolometric flux received by each planet. We present the flux-radius (F-R) relationships predicted by our framework for planets with solar metallicity envelopes. From the left to right panels, we explore low (3$M_\oplus$), intermediate (6$M_\oplus$) and high (12$M_\oplus$) mass planets. For comparison, we include observed planets with measured masses smaller than the core mass considered in each panel. Because lower-mass planets are more susceptible to atmospheric escape and therefore tend to retain smaller envelopes and radii, the observed population is expected to lie predominantly below the corresponding model F-R curves. To ensure consistency with the model assumptions, we restrict the observational sample to planets orbiting sun-like stars.

As in the M-R relationship, each F-R curve consists of two segments: a slope at high fluxes (short periods) set by escape (primarily boil-off), and a plateau at lower fluxes (long periods), where the radii reflect the initial envelope mass fractions estabished during formation. The escape-dominated slopes reproduce the observed population well for intermediate- and high-mass planets (middle and right panels), with higher-flux planets consistent with bare-core super-Earths (black curves). At lower masses, however, the trend is less apparent because of the limited number of observed systems. Moreover, the formation-determined plateaus (radius cliffs) predicted by the partially soluble models occur at radii larger than those occupied by most observed planets. The observed population is more consistent with the predictions of the fully miscible models.

Consistent with the M-R curves in Figure \ref{MR}, the F-R curves show little distinction between low-to-intermediate solubility models (1-3 times the default) and no-dissolution models. In contrast, the fully miscible models predict substantially smaller radii at low fluxes and larger radii at high fluxes. In addition, the fully miscible models extend the super-Earth transformation boundary more strongly than in the M-R case. They also exhibit a radius peak associated with outgassing, although this feature becomes less prominent at higher planetary masses. However, no comparable peak is evident in the observed exoplanet population.

\subsubsection{Stellar Mass} \label{subsubsec:stellar}
\begin{figure*}
\centering
\includegraphics[width=0.9\textwidth]{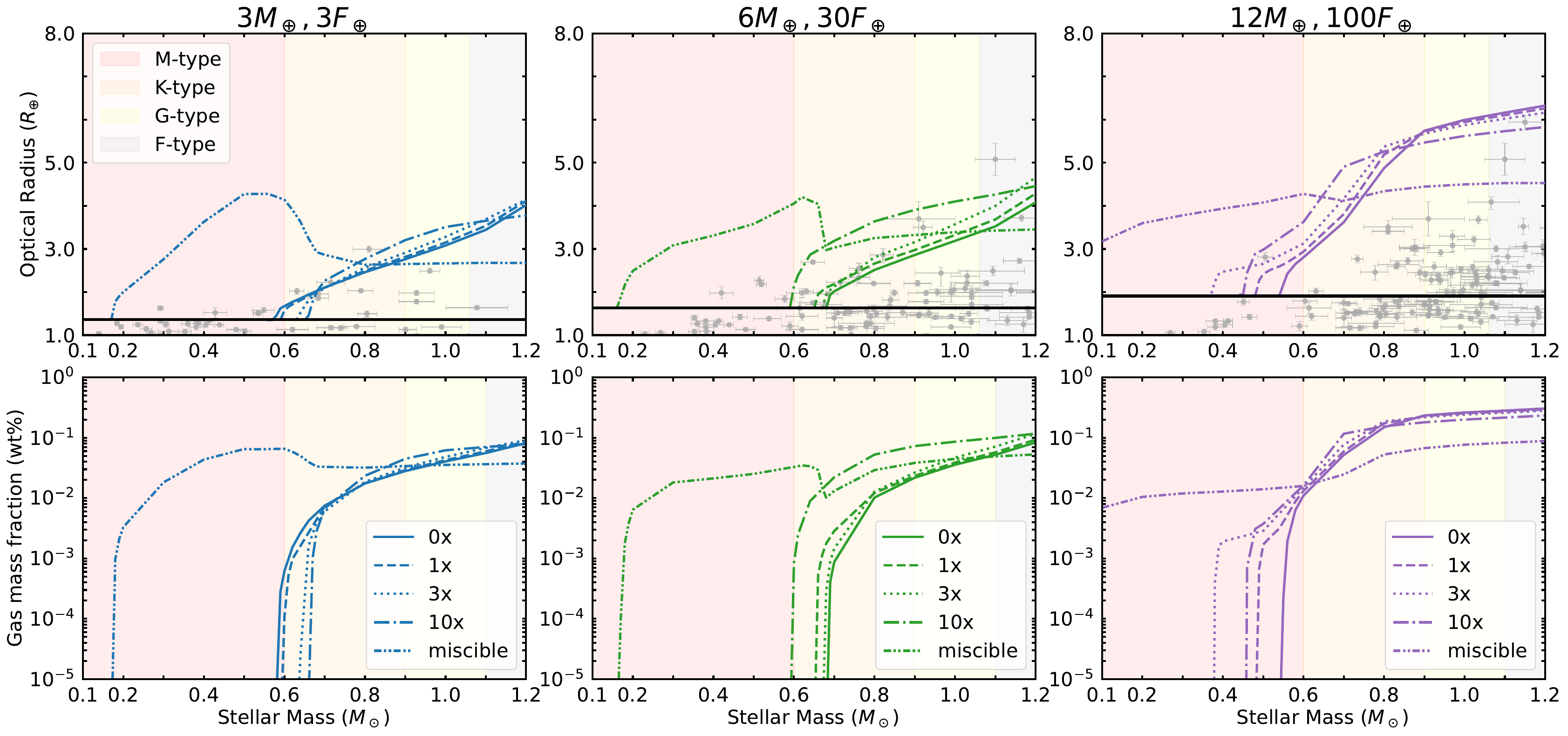} 
\caption{ Stellar-mass-radius (Ms-Rp) relationships (top row) and the corresponding gas mass fractions (bottom row). The setup is analogous to that in Figures \ref{MR} and \ref{PR}. We include all planets with well-determined stellar masses. Each panel shows Ms-Rp curves for different combinations of core mass and incident flux: 3$M_\oplus$+3$F_\oplus$ (left), 6$M_\oplus$+30$F_\oplus$ (middle), and 12$M_\oplus$+100$F_\oplus$ (right). For partially soluble models, a clear division is seen between escape-dominated segments at low stellar masses and formation-dominated segments at higher stellar masses in the right panel. We color code the background to indicate the stellar types. In each panel, we also plot all confirmed exoplanets with measured masses and incident fluxes below the corresponding core mass and flux values shown in that panel for comparison. All models have solar metalicities. See main text for discussion.
}
\label{sMR}
\end{figure*}

Figure \ref{sMR} shows the model-predicted stellar-mass-planetary-radius (Ms-Rp) relationships. In each panel, we fix the bolometric flux received at 5 Gyr to 3 $F_\oplus$ (left), 30 $F_\oplus$ (middle), and 100 $F_\oplus$ (right). The corresponding core masses are set to 3 $M_\oplus$, 6 $M_\oplus$, and 12 $M_\oplus$, respectively. We vary the stellar mass from 0.1-1.2 $M_\odot$, spanning late M dwarfs to late F-type stars. For comparison, we include observed planets with measured masses below and incident fluxes above the corresponding core-mass and flux values adopted for Ms-Rp curves in each panel. Since lower-mass and more strongly irradiated planets lose more mass through escape and therefore retain smaller radii, the observed population is expected to occur below the corresponding Ms-Rp curves.

Lower-mass stars emit a higher fraction of their luminosity in the XUV, which strengthens photoevaporation. They also undergo a prolonged pre-main-sequence contraction phase, exceeding 100 Myr for an M dwarf, during which the equilibrium temperature remains significantly elevated compared to the present-day value. This contrasts with G-type stars, whose $T_{\rm eq}$ varies only mildly after disk dispersal. The higher early-time irradiation around M dwarfs greatly enhances boil-off mass loss. As a result, both the partially soluble and insoluble models show a monotonic decrease in planetary radius toward lower stellar masses. For these models, sub-Neptunes struggle to survive around late M dwarfs unless their cores are sufficiently massive ($\gtrsim 10M_\oplus$) or they orbit at large separations (beyond a few AU). 

In contrast, the fully miscible models predict a substantially larger surviving population around M dwarfs. This arises from their strong resistance to early boil-off, which shifts the super-Earth transformation boundary to much lower stellar masses, often down to $\sim$0.2 $M_\odot$. This behavior differs from the trends seen in the M-R and F-R relationships. The characteristic outgassing-induced peak is also broader and more diffuse in stellar-mass space. As in the M-R and F-R cases, this peak weakens at high flux or large core mass.

Remarkably, in the top panels of Figure \ref{sMR}, both the predominance of super-Earths around low-mass stars and the observed slope of the sub-Neptune population are well reproduced by the partially soluble models. However, the radius cutoff, implied by the formation-determined initial envelope masses (see Section \ref{subsec:mass}), occurs at larger radii than observed. In this respect, the observed population is more consistent with the predictions of the fully miscible models. In contrast, the data do not show the pronounced radius peak near 0.5-0.6 $M_\odot$ predicted by the fully miscible models, nor do they reflect the high sub-Neptune occurrence rates around M dwarfs implied by our calculations. 

Although these results may appear to favor partial H solubility in sub-Neptune rock/iron interiors, the apparent discrepancy is likely influenced by selection biases: our sample is dominated by Kepler targets, which primarily include G-type host stars. The true sub-Neptune occurrence and radius distribution around M dwarfs therefore remain poorly constrained. These populations provide a valuable observational test of whether sub-Neptune interiors are governed by fully miscible hydrogen. Future TESS discoveries, particularly around low-mass stars, will be essential for evaluating this scenario in the future.

\subsection{Impact of Solubility Strength on Population Distribution} \label{subsec:population}
We examine the planet occurrence distribution after 5 Gyr of evolution and its dependence on solubility levels using our model results. Following the discussion in Section \ref{subsec:radius}, planetary mass, orbital period, and stellar mass all influence this distribution. Therefore, we assess each parameter independently in its respective parameter space, rather than generating a full synthetic population for direct comparison with observations.

We randomly draw 10000 planets following the distributions of these three parameters fitted from the observed samples (gray dots) in Figures \ref{MR}, \ref{PR}, and \ref{sMR}. Unlike previous models \citep{Owen17} that fit the core mass, we fit the total planetary mass, as core mass alone does not capture interior compositional information, including the envelope mass fraction and its metallicity. This leads to a Rayleigh distribution for the total mass:
\begin{equation} \label{rayleigh}
    f(M_p) = \frac{M_p}{\sigma_M^2} \exp{\left(-\frac{M_p^2}{2\sigma_M^2}\right)}
\end{equation}
where $\sigma_M$ is determined to 5$M_\oplus$. For the orbital period, we fit its corresponding logarithmic bolometric flux, $\log{F}$, which follows a normal distribution:
\begin{equation}
    f(\log{F}) = \frac{1}{\sqrt{2\pi\sigma_F^2}} \exp{\left[-\frac{{(\log{F}-\mu_F)}^2}{2\sigma_F^2}\right]}
\end{equation}
where the mean value $\mu_F=2.0$ and the standard deviation $\sigma_F=0.8$. For the stellar mass, we adopt a normal distribution peaked at 0.93$M_\odot$ with a standard deviation of 0.18, consistent with the Kepler target population. For lower-mass stars ($\leq0.6M_\odot$) observed primarily by TESS, we instead use a linear probability density function to better represent the observed distribution.

\begin{figure*}
\centering
\includegraphics[width=0.9\textwidth]{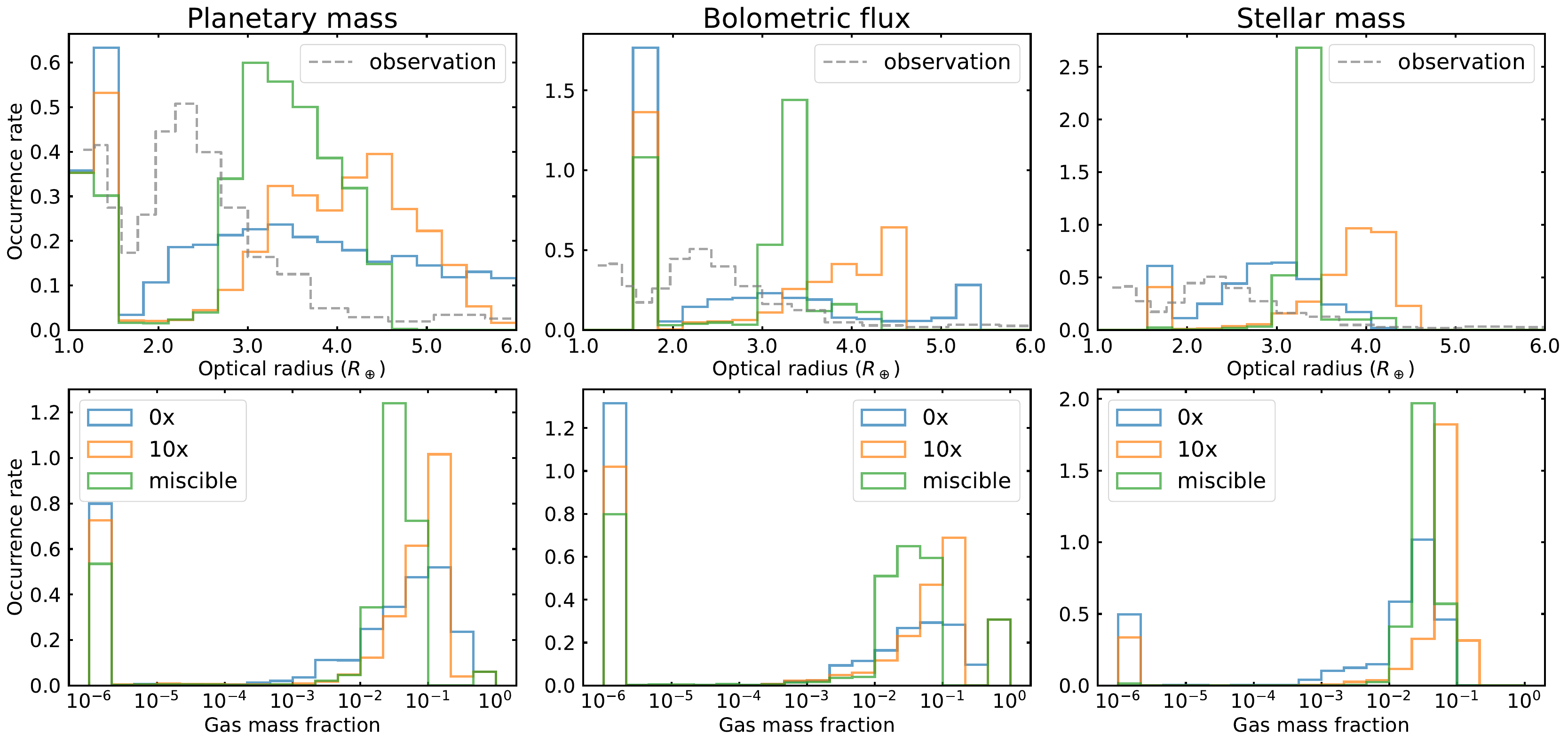} 
\caption{ Modeled population distribution for 10,000 planets. The population distribution is decomposed into three one-dimensional parameter spaces: planetary mass (left), bolometric flux/orbital period (middle), and stellar mass (right). We vary one parameter and fix the others in each column. We fit and adopt the corresponding parameter distributions from the observed sample shown in Figure \ref{MR} (planetary mass), Figure \ref{PR} (orbital period), and Figure \ref{sMR} (stellar mass). We assume solar metallicity in all panels. In each panel, we vary the solubility strength between the no-dissolution (blue), ten-times-default solubility (orange), and fully miscible (green) cases. As a comparison, the observed distribution from \citet{Fulton17} is shown in black dashed curves. The top row shows planetary occurrence rate as a function of optical radius, while the bottom row shows the occurrence rate as a function of gas mass fraction. See the main text for further discussion.
}
\label{popSol}
\end{figure*}

Figure \ref{popSol} shows radius (top) and corresponding gas mass fraction (bottom) distributions in the planetary mass, bolometric flux (orbital period) and stellar mass spaces in the left, middle and right panels. The default parameters correspond to a core mass of 6$M_\oplus$, irradiation of 30$F_\oplus$, and a Sun-like host star, representing the most commonly observed sub-Neptune configuration (same as the second-from-right panels of Figure \ref{MR}, \ref{PR} and \ref{sMR}). In each panel, we show the population distributions for the no-dissolution (blue), 10x (orange), and fully miscible (green) models. The 1x and 3x solubility cases are not shown, as their population distributions are effectively indistinguishable from the no-dissolution model.

In the planetary mass and bolometric flux spaces, increasing the solubility level reduces the radius cutoff and significantly widens the gap. In contrast, in the stellar mass space, the no-dissolution model (blue) exhibits the smallest radius cutoff, comparable to that of the fully miscible model whereas the 10x solubility model shows the highest cutoff radius. Here, the fully miscible model does not produce a radius gap because these planets are largely insensitive to boil-off (see Section \ref{subsubsec:stellar}). The occurrence rate of super-Earths decreases with increasing solubility level, while the fully miscible model predicts the highest peak in the sub-Neptune population and an overabundance of sub-Neptunes relative to super-Earths.

Among the three parameters, planetary mass most strongly acts to widen the sub-Neptune distribution, whereas stellar mass tends to compress it substantially, especially in the no-dissolution case. In contrast, bolometric flux has a mixed effect that depends on the solubility level. Understanding how these influences combine to shape the observed radius gap and radius cutoff will require additional, dedicated population-modeling efforts.

\subsection{Role of Metallicity} \label{subsec:metal}
Metallicity influences radius evolution through several competing effects. Higher metallicity increases atmospheric opacity, causing planets to cool more slowly and thus inflating their radii. This delayed cooling also suppresses outgassing. At the same time, metals increase the interior density and reduce the atmospheric scale height, both of which act to shrink the radius. On the other hand, higher metallicity decreases the optical transit pressure, buffering the reduced planetary radius; this effect is small due to the small atmospheric scale height. The reduced scale height also weakens atmospheric escape, which does not necessarily translate to a smaller radius at a given evolutionary age. These complicated interactions necessitate a coupled interior-composition-escape model like SEAMIST.

\begin{figure}
\centering
\includegraphics[width=0.5\textwidth]{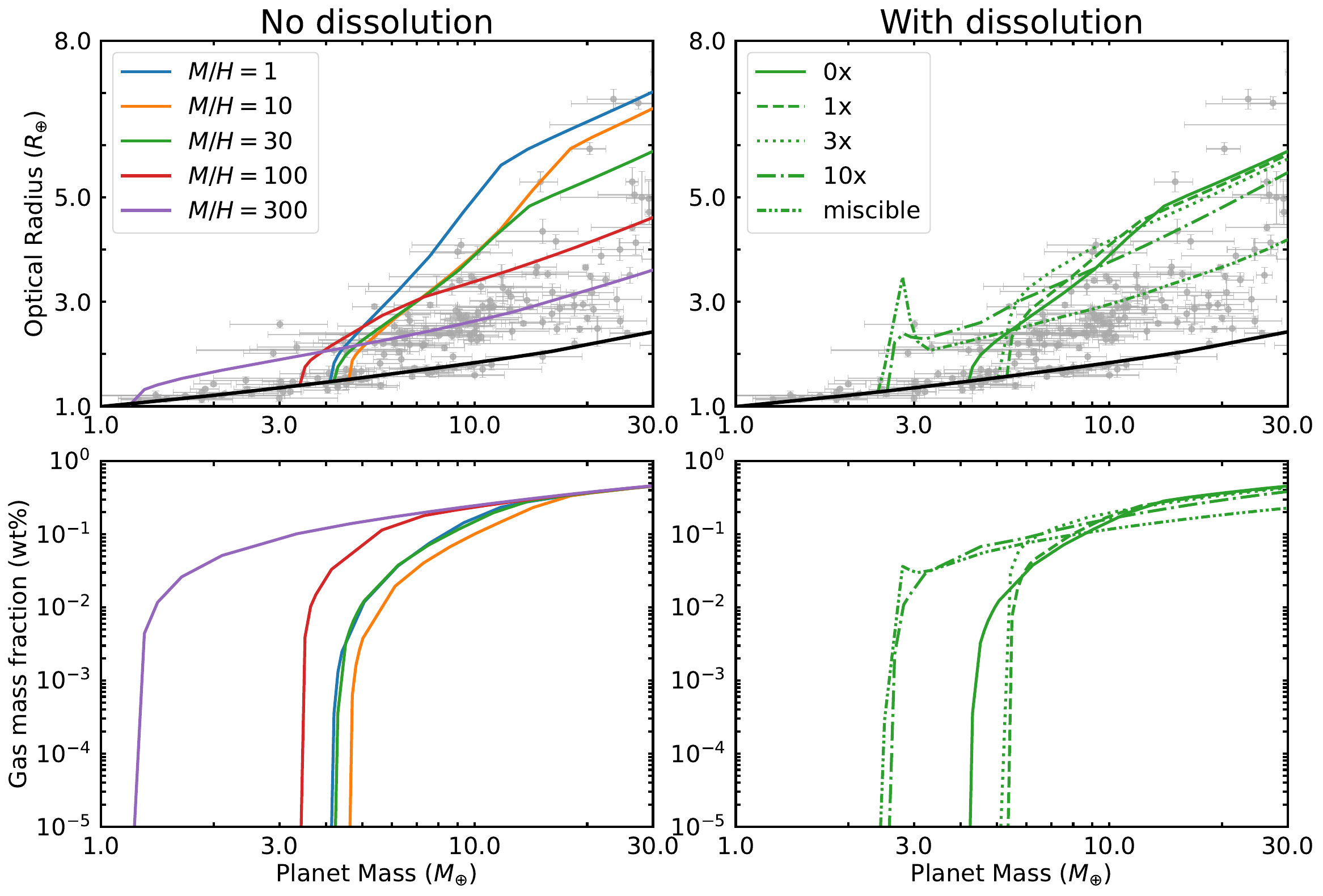} 
\caption{ The left column shows the M-R relations (top) and their corresponding gas mass fraction curves (bottom) for metallicities ranging from 1 to 300 times solar, assuming no dissolution. The right column illustrates the impact of volatile solubility on intermediate-metallicity planets (30× solar), using a setup analogous to that in Figure \ref{MR}.
}
\label{MetR}
\end{figure}

Given these coupled influences, especially in combination with volatile dissolution, we compute a grid of models to explore how the M-R relationship depends on metallicity. The results are summarized in Figure \ref{MetR}. In the left panels, we show our baseline models with metallicity varying from 1 to 300 times solar, with dissolution turned off. For planets with masses $\gtrsim 10M_\oplus$, the radius decreases monotonically with increasing metallicity (top). At lower masses, however, the trend reverses once the metallicity exceeds $\sim$30x solar, extending the super-Earth transformation boundary to higher masses. This extension is limited except for the highest metallicities, which predict a sub-Neptune population at $1M_\oplus$. 

The radius cutoff associated with the observed overdensity below $4R_\oplus$ is best reproduced by the 100x solar models for planets with masses $\gtrsim10M_\oplus$. In contrast, the lower-metallicity curves provide a better match to the small number of rare, large-radius planets that lie above the main population. At lower masses, however, the data remain consistent with the escape-dominated segments of all models with metallicities $\leq100$x solar. 

\begin{figure}
\centering
\includegraphics[width=0.5\textwidth]{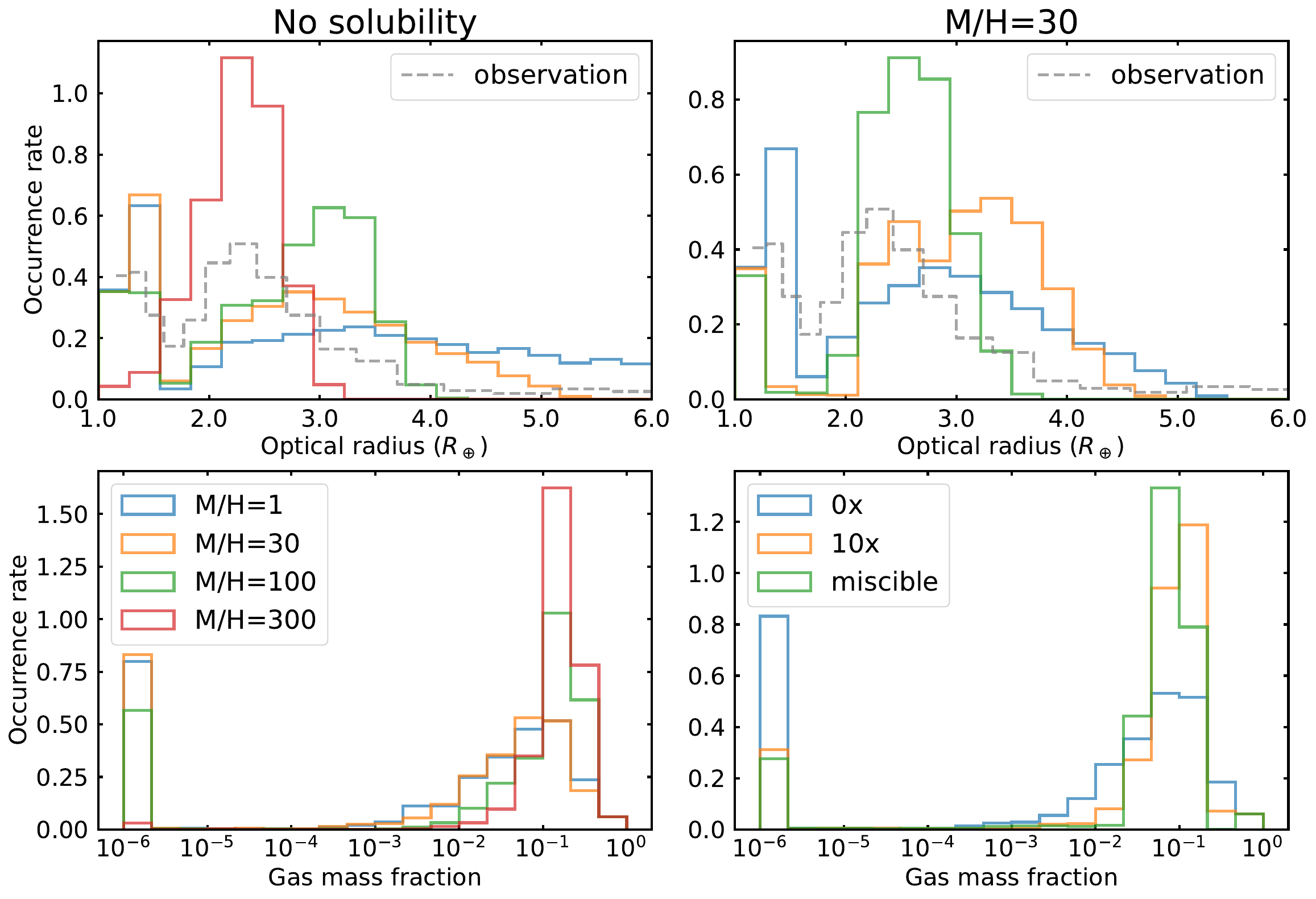} 
\caption{ The corresponding population distributions for the M-R curves shown in the top panels of Figure \ref{MetR}. The core mass distribution follows Eq. \ref{rayleigh}, consistent with that used in Figure \ref{popSol}.
}
\label{popMet}
\end{figure}

These behaviors are reflected in the modeled population distributions. In the left panels of Figure \ref{popMet}, we show population results for the 1x (blue), 30x (orange), 100x (green), and 300x (red) models using the same mass distribution as in Section \ref{subsec:population}. With increasing metallicity, the radius cutoff shifts downward and the sub-Neptune distribution becomes more concentrated. The location and depth of the radius valley remain nearly unchanged for metallicities $\leq$100x solar, whereas the 300x models produce no radius gap at all due to the limited mass removal (bottom).

We fix the metallicity to 30x solar and vary the solubility strength in the right panels of Figure \ref{MetR}. The resulting changes in the M-R curves are substantially more pronounced than in the low-metallicity case. At 1-3x the default solubility, the super-Earth transformation boundary actually shifts to larger masses relative to the no-dissolution model, indicating that outgassing strengthens escape in this regime. Similar to Figure \ref{MR}, the differences among the curves at higher masses remain minor. However, at 10x the default solubility, the transformation boundary abruptly moves to much lower masses, revealing that outgassing-driven mass replenishment has become the dominant process. In this regime, outgassing remains dominant until atmospheric escape rapidly takes over, producing a sharp downturn at the boundary, an effect closely analogous to that seen in the fully miscible models. Consequently, the 10x-solubility curve resembles the fully miscible case at low masses, except for the pointy peak.

In the right panels of Figure \ref{popMet}, we vary the solubility levels. Similar to Figure \ref{popSol}, the 1x and 3x models show no significant deviation from the no-dissolution case. As expected, the radius cutoff shifts to smaller radii with increasing solubility, consistent with the behavior seen in the solar-metallicity case (Section \ref{subsec:population}). However, in contrast to the solar-metallicity trends, we find no evidence of an additional widening of the sub-Neptune distribution between the 10x and fully miscible models, and even the widening between the no-dissolution and 10x cases remains limited.

\section{Discussion} \label{sec:disc}
\subsection{Implications for a Sub-Neptune Origin of Earth}
Previous studies have suggested the potential existence of a low-metallicity primordial envelope on the young Earth \citep{Owen:review2020,Young23}.  To evaluate this possibility, we perform coupled evolution calculations to determine if such an envelope could be removed by atmospheric escape, producing the envelope-free Earth we see today.

We run models spanning the full range from no dissolution to partially soluble to fully miscible envelopes. In all cases, we find that if the initial metallicity is low, Earth's primordial H/He envelope, whether initially massive (5\%, as predicted by GSS16) or thin (0.2\% per \citet{Young23}), would be lost within $\leq$100 Myr. After this, a thin primordial radiative atmosphere could persist for an additional $\sim$100 Myr, if no secondary atmosphere forms. These results indicate that, within our framework, a sub-Neptune origin for Earth cannot be ruled out.

\subsection{Hydrogen in Magma Ocean: Partial Soluble or Fully Miscible?}
By comparing our model predictions with the observed exoplanet population across multiple parameter spaces, we identify several observational signatures that can be used to test H solubility in sub-Neptune interiors. Our results show that the partially soluble models more successfully reproduce both the atmospheric-escape-driven slopes of the sub-Neptune population and the location of the transition to the super-Earth population. However, these models struggle to explain the observed radius cliff--the sharp decline in planet occurrence near $3.5,R_\oplus$--among sub-Neptunes with massive envelopes that are largely resistant to atmospheric escape. In contrast, the fully miscible models reproduce the observed radius cliff remarkably well, but appear less consistent with the observed distribution of smaller sub-Neptunes that are strongly affected by atmospheric escape, particularly in stellar-mass space. %Taken together, the current observations provide modest support for partially soluble hydrogen over the fully miscible scenario, although the latter cannot be ruled out. 
In the following sections, we discuss several possible physical explanations for the discrepancies between the model predictions and the observations.

\subsection{Radius Peak in Fully Miscible Models}
None of Figure \ref{MR} or Figure \ref{sMR} shows evidence for the predicted radius peak (the dash-dot-dot curves) in the fully miscible models. Several factors may explain this discrepancy: (1) large uncertainties in mass measurements for low-mass planets; (2) the limited sensitivity of the Kepler mission to low-mass stars, which may undersample the relevant population, an issue that ongoing TESS discoveries of late-type systems may help resolve; and
(3) the omission of chemical reactions and metal outgassing in our model. For example, if dissolved H reacts with FeO in the mantle to form $\rm H_2O$, subsequent outgassing of water would raise the atmospheric mean molecular weight, counteracting H outgassing and potentially altering the physics that sets the radius peak. However, the amount of H outgased as water sensitively depends on the mantle's redox state and chemical equilibrium \citep{Kite21,Lichtenberg21,Krissansen-totton24}, and thus requires future investigation. (4) when the MEB temperature falls below the binodal temperature (corresponding to planets with cooler interiors), hydrogen may still remain highly soluble. Accurately capturing the transition between the fully miscible and partially soluble regimes is therefore essential for robust predictions.

Improved observational samples, together with chemical modeling, will provide stronger constraints on the role of H-rock-iron miscibility in shaping the observed population distribution.

%we propose that this radius peak could serve as an observational test of H-rock-iron miscibility.

\subsection{Implications for the Radius Cliff}
Our results show that the radius cliff is primarily set by the initial conditions inherited from planet formation, and is comparatively insensitive to atmospheric escape. At solar metallicity, our partially soluble model predicts a significantly higher radius cutoff than observed (Figures \ref{MR}, \ref{PR}, and \ref{popSol}). Several possibilities may explain this discrepancy:
\begin{enumerate}
    \item Smaller primordial envelopes: Sub-Neptunes may commonly form with a more limited initial gas inventory, consistent with late-stage core assembly \citep{Lee14}, rather than with the more massive envelopes predicted by GSS16. \footnote[1]{GSS16 prescription focuses on gas accretion, which is appropriate for the low-metallicity regime. In the high-metallicity regime, the metal volatile inventories are expected to be limited by solid accretion beyond the snowline rather than by gas accretion, and thus our adoption of the GSS16 prescription may shift the initial envelope mass, impacting the radius cliff.}
    \item Additional mass-loss mechanisms: Processes operating where both photoevaporation and boil-off are inefficient (e.g., high mass, low flux) could further reduce radii. Giant impacts can remove additional envelope mass through both shock heating of the envelope, which enhances hydrodynamic escape \citep{Inamdar16}, and direct kinetic stripping via momentum transfer and shock-driven ejection \citep{Kegerreis2020} and therefore represent a plausible candidate. However, invoking them would disrupt the close agreement between our escape-driven M-R sequence and the observed population.
    \item Early-time internal heating: Young planets may experience additional luminosity from tidal dissipation or rock/iron differentiation, which could drive further mass loss. However, this does not explain the smaller observed radius cutoff at wide separations (Figure \ref{PR}), nor the fact that low-mass planets, expected to be more sensitive to core heating, do not show the corresponding mismatch (Figure \ref{MR}).
    \item Fully miscible H-rock behavior: Complete miscibility and the resulting He-rich planets can reproduce the smaller observed cutoff, but fails to account for the substantial population of outliers. Current models and observational data disagree on the location of the predicted radius peak.
    \item High atmospheric metallicities: Elevated metallicity effectively reduces planetary radii and can also accommodate the observed outliers, making it a promising explanation. The small radius cutoff is likely to reflect metal-rich worlds (see Figure \ref{MetR}).
\end{enumerate}
In summary, the physical origin of the radius cutoff remains uncertain; it could result from both full miscibility and high metallicity. Based on our modeling, we slightly favor scenarios in which the smaller observed cutoff arises primarily from a high atmospheric metallicity, potentially combined with a gas-poor ($\lesssim$ a few percent) formation channel and additional, non-radiation-driven atmospheric escape.

\subsection{Other Limitations} \label{disc:limitation}
We neglect changes in the thermodynamic properties of the mantle and core, such as viscosity, thermal conductivity, thermal diffusivity, heat capacity, and melting temperature, that would result from evolving volatile content. In reality, H-silicate and H-Fe mixtures will have different cooling and solidification behaviors as the dissolved mass fraction evolves. Likewise, Eqs. \ref{silicateevolution} and \ref{ironevolution} do not include compositional cooling/heating terms for the rock/iron core. Because volatiles carry distinct specific entropies compared to silicate and iron, changes in mixing ratio during outgassing should alter the potential temperatures $T_m$ and $T_c$. These changes would enter the thermal budget through terms analogous to $ds/dt_o$ and $ds/dt_m$ in Eq. \ref{thermalenvelope}. However, existing high-pressure experiments and first-principle simulations are currently unable to robustly constrain the relevant thermodynamic properties of H-silicate and H-Fe mixtures. Addressing this limitation will require improved constraints on these mixture properties. 

We also neglect tidal heating and coupled orbital dynamics, as is commonly done in recent sub-Neptune models \citep{Aguichine25,Rogers25}. Tidal dissipation and high-eccentricity migration can potentially influence volatile outgassing by heating the rock/iron interior, while also enhancing atmospheric mass loss through an increase in the atmospheric scale height. In addition, giant impacts may further reduce the retained envelope mass fraction, thereby decreasing the planetary radius. Incorporating these effects within a unified evolutionary framework remains an important direction for future work. In the context of impact erosion, Figures \ref{MR}, \ref{PR}, \ref{sMR}, and \ref{MetR} should be interpreted as providing upper-limit predictions for planetary radii and envelope mass fractions. These predictions nevertheless offer meaningful theoretical constraints for comparison with the observed exoplanet population.

\section{Conclusions} 
In this work, we present SEAMIST, a sub-Neptune evolution framework that, for the first time, self-consistently couples thermal evolution, boil-off, photoevaporation, rock/iron solidification, H/He dissolution, and atmospheric compositional evolution within a single model. 

Previous planetary evolution studies have often treated the initial envelope mass fraction as a free parameter, tuning it to reproduce the observed radii of individual planets and population-level distributions. Although this approach has successfully matched observations, the adopted initial conditions vary widely across models and depend strongly on the included physics, leading to substantial degeneracies in the inferred mechanisms. These degeneracies are further amplified in population synthesis studies by assumptions about underlying distributions, such as the core mass function: by adjusting these inputs, fundamentally different mass-loss prescriptions can reproduce similar observed populations, limiting their physical interpretability and their ability to explain multiple observational trends simultaneously.

To overcome these limitations, we unify the relevant physical processes and evolve planets from self-consistent initial conditions informed by planet formation models. We then follow their coupled evolution to 5 Gyr, enabling a more physically grounded assessment of how sub-Neptune radii and populations emerge across a high-dimensional parameter space. 

Using SEAMIST, we make theoretical predictions for how planetary mass, orbital period, stellar mass, stellar metallicity, and volatile solubility strength influence both final planetary radii and population-level distributions. We also identify several new physical mechanisms that govern the coupled thermal and atmospheric evolution, improving people's existing understanding. By comparing our results with observations, our framework provides new insights into the demographics of small exoplanets. The key takeaways are:
\begin{itemize}
    \item We find that the inclusion of volatile solubility does not necessarily prolong the envelope lifetime. This result contrasts with previous work by \citet{Chachan18}, which suggested that dissolved volatiles act as a replenishing reservoir that buffers atmospheric mass loss. The discrepancy arises from our self-consistent treatment of atmospheric compositional evolution and its impact on interior evolution. In our models, hydrogen is outgassed from the silicate-iron interior more efficiently than helium, leading to an increase in the hydrogen abundance over time. This lowers the mean molecular weight of the envelope, reduces its density, increases its specific entropy, and expands the atmospheric scale height, thereby enhancing mass-loss rates for both photoevaporation and boil-off.
    \item At high hydrogen solubility, a sub-Neptune's envelope is often He-dominated or metal-dominated. This provides an alternative pathway to forming He-rich worlds, in addition to atmospheric fractionation during atmospheric escape \citep{Cherubim24}, which occurs primarily in high-gravity outflows. As atmospheric escape rapidly reduces the envelope mass fraction to a few percent, the remaining envelope gradually becomes H-dominated due to the H replenishment. At later stages, however, the mean molecular weight may increase again due to helium outgassing when the mass fraction drops to $\sim$0.1\%.
    \item Our predicted M-R curves at solar metallicity define a theoretical upper limit on planetary radii and show strong agreement with the observations. At low planetary masses, the M-R relation is tightly constrained by atmospheric escape, and this escape-dominated regime closely matches the observed M-R distribution. At higher masses, however, the observed radius cut-off lies substantially below our theoretical upper envelope, indicating that processes other than atmospheric escape dominate the formation of smaller radii in this regime. We argue that this discrepancy points to one or more of the following possibilities: systematically lower initial envelope mass fractions than predicted by formation models such as \citet{Ginzburg16}, higher atmospheric metallicities, additional mass loss from giant impacts, or efficient, fully miscible H-silicate interactions in the planetary interior.
    \item We identify three evolutionary regimes in the fully miscible models: (i) a cooling-dominated regime at high masses, in which slow cooling leads to near-complete dissolution of H in the interior and a He-dominated envelope; (ii) an outgassing-dominated regime at lower masses, in which rapid H exsolution inflates the planetary radius over time; and (iii) an escape-dominated regime at the lowest masses, in which atmospheric escape efficiently removes the outgassed material through either boil-off or photoevaporation. This produces a radius peak at low planetary masses that is not seen in current observations. We therefore propose that the presence or absence of this feature can serve as an observational test of hydrogen solubility/miscibility in sub-Neptune interiors in future surveys.
    \item Our model predicts that sub-Neptunes are unlikely to survive around M dwarfs unless their envelopes are highly metal rich or hydrogen is fully miscible with silicates. If such planets do exist, we would expect the radius gap to be absent for this population (see Figure \ref{popMet}), which has been suggested by recent observation \citep{Gillis26}.Future TESS observations can test this prediction and thereby provide an empirical constraint on hydrogen miscibility inferred from first-principles simulations. These observations will also provide a critical testbed for our coupled evolution model.
    \item At high solubility levels, we find that sub-Neptune envelopes are primarily composed of helium rather than hydrogen, with metals becoming increasingly important at high metallicity. This naturally helps explain the observed radius cutoff: as hydrogen solubility increases, the cutoff shifts toward smaller radii. However, this process can also lead to a widening of the radius gap. The broader implications of this effect will need to be explored in future work.
    \item We identify a catastrophic boil-off driven by a positive feedback between atmospheric mass loss and hydrogen outgassing, mediated by the increasing hydrogen abundance in the envelope. This feedback produces an extraordinarily short mass-loss timescale of order $10^3$ yr. Remarkably, it allows boil-off to be triggered at late evolutionary stages, from millions to billions of years after disk dispersal.
\end{itemize}

\section{Acknowledgments} 
We thank the anonymous reviewer for valuable comments
that improved this work. We also thank James Rogers and Hilke Schlichting for discussions about hydrogen miscibility, Yayaati Chachan for valuable insights into hydrogen dissolution. and Artem Aguichine for providing the steam-atmosphere models. J.J.F. acknowledges the support of NASA's Interdisciplinary Consortia for Astrobiology Research
(NNH19ZDA001N-ICAR) under award number 19-ICAR19
2-0041.

\bibliography{reference}{}
\bibliographystyle{aasjournal}

%% This command is needed to show the entire author+affiliation list when
%% the collaboration and author truncation commands are used.  It has to
%% go at the end of the manuscript.
%\allauthors

%% Include this line if you are using the \added, \replaced, \deleted
%% commands to see a summary list of all changes at the end of the article.
%\listofchanges

\end{document}